\documentclass[9pt,twocolumn,twoside]{pnas-new}
\setboolean{displaywatermark}{false}
\usepackage{multicol}
\usepackage{subfig}

\usepackage{hyperref,comment} 

\title{Torsional Instabilities in the Delamination of Soft Adhesives}

\author[a,b]{Tara K. Venkatadri}
\author[c]{Thomas Henzel}
\author[*,c,d]{T. Cohen}

\leadauthor{Venkatadri} 

\affil[a]{Department of Aeronautics and Astronautics, Massachusetts Institute of Technology, Cambridge, MA 02139}
\affil[b]{Department of Engineering, University of Cambridge, Trumpington Street, Cambridge, CB2 1PZ, UK}
\affil[c]{Department of Civil and Environmental Engineering, Massachusetts Institute of Technology, Cambridge, MA 02139}
\affil[d]{Department of Mechanical Engineering, Massachusetts Institute of Technology, Cambridge, MA 02139}

\begin{abstract} 
Soft adhesive contacts are ubiquitous in nature and are increasingly used in synthetic systems, such as flexible electronics and soft robots, due to their advantages over traditional joining techniques. While methods to study the failure of adhesives typically apply tensile loads to the adhesive joint, less is known about the performance of soft adhesives under shear and torsion, which may become important in engineering applications. A major challenge that has hindered the characterization of shear/torsion-induced delamination is imposed by the fact that, even after delamination, contact with the substrate is maintained, thus allowing for frictional sliding and re-adhesion. In this work, we address this gap by studying the controlled delamination of soft cylinders under combined compression and torsion. Our experimental observations expose the nucleation of delamination at an imperfection and its propagation along the circumference of the cylinder. The observed sequence of `stick-slip' events and the sensitivity of the delamination process to material parameters are explained by a theoretical model that captures axisymmetric delamination patterns, along with the subsequent frictional sliding and re-adhesion. By opening up an avenue for improved characterization of adhesive failure, our experimental approach and theoretical framework can guide the design of adhesives in future applications. 

\end{abstract}

\dates{$^*$ To whom correspondence should be addressed: T. Cohen  (talco@mit.edu)}

\begin{document}

\maketitle

\noindent 
The advantages of adhesives, in comparison with traditional joining techniques, have fueled their development in various industries, with emerging applications in automotives \citep{alfano2018fracture}, aerospace \citep{kinloch1997adhesives,davies2006predicting,barzegar2019numerical}, structural engineering \citep{mays1992adhesives,buyukozturk1998failure,rahimi2001concrete,buyukozturk2004progress,hartshorn2012structural} medical practice \citep{li2016bioinspired}, and robotics \citep{jiang2017robotic}. In contrast to mechanical fastening and welding, adhesives allow for the joining of dissimilar materials, have improved fatigue performance, and can reduce stress concentrations \citep{bowditch1996adhesive, shang2019strategy, banea2010effect}. Such improvements are traditionally estimated via testing in specific plane-strain geometric settings, such as `lap-joint' tests \citep{banea2010effect, bowditch1996adhesive, shang2019strategy} and peeling tests \citep{bowditch1996adhesive, cohen2018competing, collino2014detachment, ringoot2021stick}, or in axially symmetric settings, such as `probe tack tests' using either flat or spherical punches \citep{shull1998axisymmetric}. However, these results cannot be directly translated to explain and predict the performance of adhesive joints in more complex loading conditions. 

We can classify the failure of adhesives into two broad categories \citep{oh2007strength}: \textit{(i)} failure in the bulk of the adhesive, and \textit{(ii)} failure at the interface between the adhesive layer and the materials that it is bonding, also known as peeling or delamination. While several studies have been devoted to enhancing the durability of adhesives and preventing delamination by tuning the material composition \citep{bowditch1996adhesive,liu2018comparison,khashaba2020development, razavi2017mixed}, the loading state and geometry of the system can also affect the strength and failure mode of the adhesive joint \citep{bartlett2012looking}. For example, peeling tests are commonly conducted by pulling the adhesive off the substrate at an angle, and the force at which delamination occurs has been shown to depend on the loading angle \citep{collino2014detachment}. Additionally, recent work on longitudinally-loaded rectangular adhesive pads \citep{cohen2018competing, ringoot2021stick} discovered a tradeoff between two failure modes (cavitation and curling) that depends on the dimensions of the adhesive and its ability to re-establish the bond upon contact with the substrate after initial peeling. 

One phenomenon that has garnered particular interest is the delamination of adhesive joints subjected to torsional loading. Several researchers have conducted torsion-delamination tests with rigid spherical indenters in contact with a planar adhesive sample \citep{chateauminois2010friction, hetenyi1958contact}, while others have studied tubular joints where the adhesive layer is sandwiched between two coaxial cylindrical shells \citep{bryant1965measurement, oh2007strength, shishesaz2019effects}. However, rather than examining the progression of delamination in detail, these analyses are geared towards quantifying the interfacial toughness and thus focus on a limited set of observations (i.e. the initiation of delamination). 
Accordingly, such experiments are configured to minimize the influence of transient interfacial phenomena. Spherical adhesive joints are used to produce localized deformations at a small point of contact between the indenter and the substrate. Similarly, tubular joints (in which adhesives bond the lateral surfaces of two coaxial cylinders) minimize the thickness of the adhesive zone and often are not conducive to visual observation, especially for joints with adherends made of an opaque material like steel \citep{bryant1965measurement}. Hence, alternative methods are needed to obtain measurements that may explain and predict delamination in realistic settings, where the adhesive bond does not `fail' instantaneously, but rather undergoes a complex delamination process that is highly dependent on the interfacial geometry and the presence of imperfections.

Our study relies on a different geometry: a cylindrical adhesive sample with one circular face in contact with a substrate. We apply axial compression to adhere the sample to the glass, then we apply torsion and observe the initiation and progression of delamination. Consistent with prior work, we have chosen to use the naturally-adhesive elastomer polydimethylsiloxane (PDMS) as our laboratory model of an adhesive material, and we have used glass as the substrate \citep{chateauminois2010friction, chaudhury2007studying}. This configuration has multiple advantages over the ones mentioned above. It provides a better model of the real-world applications of adhesives compared to the spherical indenter tests, since adhesives under torsion (e.g. composite laminates) are more likely to have adhesive contact in a planar setting than on a sphere. In addition, the use of a transparent adhesive and substrate in contact on the cylinder's circular surface (instead of the lateral surface) allows for visual tracking of the delamination process in a way that the tubular joint tests do not. This setup therefore enables us to observe behaviors beyond the initiation of delamination and to capture the effect of nonlinear deformations, which have recently been shown to play a significant role in determining the interfacial toughness of soft adhesives \citep{interfacial,wahdat2022pressurized}.

Some preliminary research has already been conducted on the cylinder/flat-plate geometry we propose. The linear elastic model created by \citet{perez2021incipient} considers the initiation of delamination, but did not capture the reattachment and stick-slip behavior. \citet{chaudhury2007studying} studied the same configuration through an experimental lens; they placed a glass disk in torsional contact with a thin PDMS film and observed the formation of delamination cavities and stick-slip behavior consistent with prior observations of Schallamach waves on adhesive joints \citep{barquins1976friction, schallamach1971does}. To minimize effects of bulk deformation, \citet{chaudhury2007studying} focused on thin films. However, in some adhesive applications (such as robotic grippers), the adhesive’s thickness may be on the same order of magnitude as its radius, so it is necessary to observe the deformation of thick adhesive layers under torsion as well. 
Our work aims to address the limitations of these prior studies, in order to develop a comprehensive understanding of the initiation and progression of delamination in cylindrical adhesive joints. Our experimental configuration also enables the investigation of stick-slip cycles and the propagation of Schallamach waves \citep{schallamach1971does}, where the adhesive-substrate interface ruptures and then re-establishes the bond \citep{ringoot2021stick}.

This manuscript is organized as follows: In the next section, we detail our experimental procedure and discuss the experimental findings. These observations agree with the key results of a minimal theory, presented in Section \ref{Theoretical Framework}, which captures both the delamination, frictional sliding, and stick-slip cycles of cylindrical adhesives under combined compression and torsion. We show results of the model in Section \ref{Results and Sensetivity Analysis} and conclude in Section \ref{section:Conclusion}.
\raggedbottom

\section{Experimental Observations} 
\label{section:ExperimentalObservations}
\begin{figure}[h]
    \centering
    \includegraphics[width=0.45\textwidth]{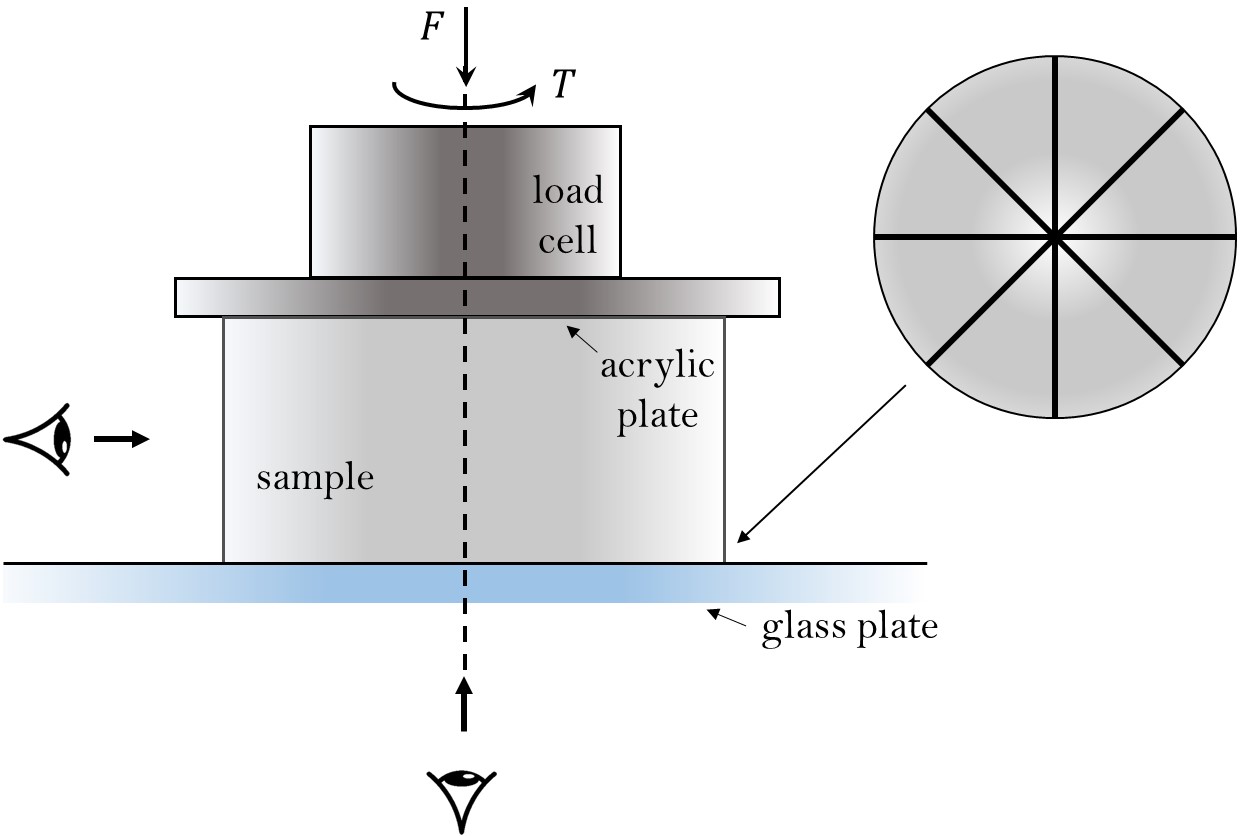} 
    \captionsetup{justification=centering}
    \caption{{Schematic illustration of the experimental configuration}. The top surface of the PDMS sample is rigidly attached to an acrylic plate that can be rotated in the plane or vertically translated, as controlled by an Instron universal testing machine. The normal force and the torque are measured by the load cell. The bottom surface of the sample is marked with radial lines (which allow for tracking of the delamination process) and put in weak adhesive contact with a glass plate. The sample is viewed both from the side and from below.
    } 
    \label{fig:MainTextSchematic}
\end{figure}

To examine the delamination patterns that emerge under a combination of compressive and torsional loading, we developed a cylindrical experimental model (Fig. \ref{fig:MainTextSchematic}). We fabricated cylindrical samples of PDMS (Sylgard 184) with diameter 2.4 cm, height 1 cm, and a base:cross-linker ratio of 40:1 to yield an elastic modulus $E \sim 25$ kPa, as measured from the torsion data in the linear range. We chose to limit our analysis to these small samples due to the high sensitivity of wider samples to imperfections in the tilt of the experimental setup (as discussed in the ESI). We placed one circular surface of each sample in weakly-bonded adhesive contact with a glass plate (see ESI for fabrication procedure). The other cylindrical surface was rigidly bonded to an acrylic disk using cyanoacrylate glue. Each acrylic disk was given a 3-digit Serial Number (SN) to aid in data processing. 
 
We attached the acrylic disk to the load cell of a dynamic Instron ElectroPuls E3000 universal testing machine and performed a combined compression-torsion test on the samples (details are given in ESI). To characterize the effect of compressive normal force on the torsional delamination process, we initially applied one of three compressive force values $F=(0, 0.5,1)$ N, then rotated the load cell at a quasi-static rate of $\dot{\alpha}=0.2$ deg/sec, up to the final rotation angle $\alpha=150$ degrees, while holding the load cell’s vertical position constant. During the rotation, we measured the variation in torque $(T)$ and normal force $(F)$. 

\begin{figure*}[!ht]
    \centering\includegraphics[width=0.95\textwidth]{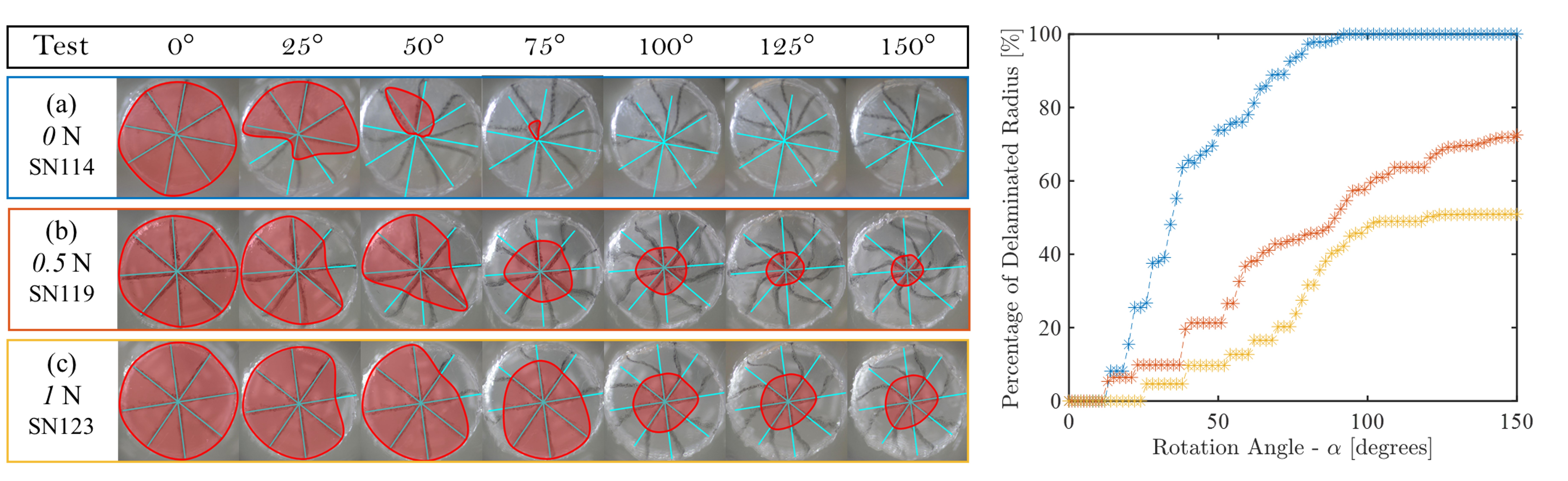} 
    \captionsetup{justification=centering}
    \caption{{Typical progression of delamination.} On the left, the sequence of images shows delamination at different rotation angles for one representative sample in each of the three categories of applied normal force. The red-shaded region indicates the area that has not yet delaminated. The radial and circumferential progression of delamination is apparent, as is the change in the final delaminated state with increasing applied compression. On the right, the corresponding evolution of the surface-averaged delaminated radius is shown. The boxes around the sets of images (on the left) correspond to the colors of the curves (on the right). Time-lapse slideshows of the delamination process for these three tests can be found at the following \href{https://drive.google.com/drive/folders/1XjKZ8HKTsY8BhSIHX8Wy9IqrMPHnHHpo?usp=sharing}{\textcolor{blue}{link}}.
    }
    \label{fig:ProgressionOfDelamination}
\end{figure*}

\begin{figure*}[!ht]
    \centering
    \includegraphics[width=1\textwidth]{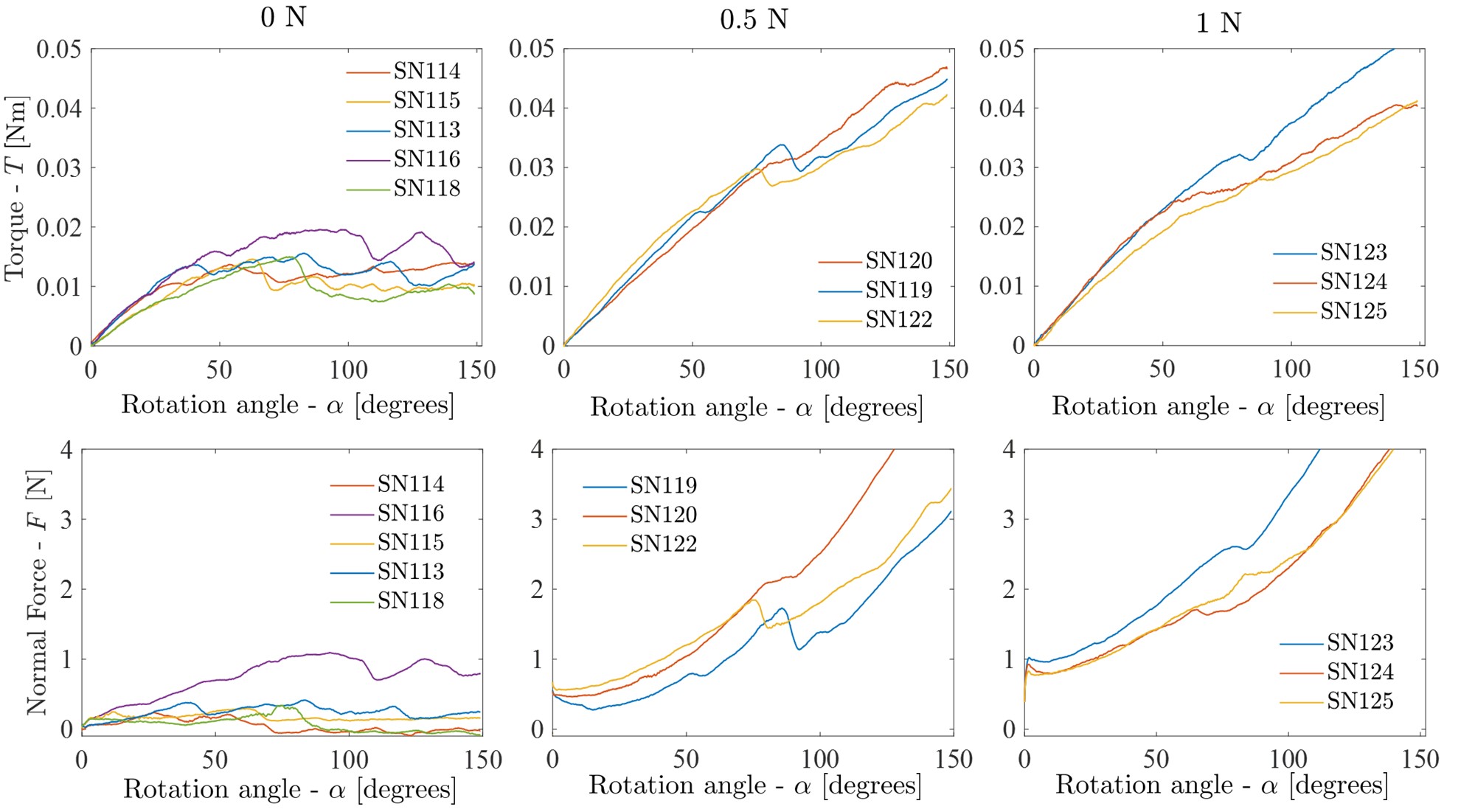} 
    \captionsetup{justification=centering}
    \caption{{Torque (top) and normal force (bottom)}, shown as a function of the rotation angle for the three initial values of normal force (columns). The variability between samples is caused by the imperfection sensitivity of the delamination process, and is most significant for samples with no initially-applied normal force. Correspondence between jumps in torque and jumps in normal force is also observed, and is associated with rapid delamination events (see ESI). The outlier tests (SN117, SN121, and SN126) are not shown on the graphs (see ESI).}
    \label{fig:TorqueAndNormalForce}
\end{figure*}

We tracked the displacement of the sample at the interface with the glass plate by marking 4 straight lines at equal angular spacing across the diameter of the PDMS surface (Fig.\ \ref{fig:MainTextSchematic}). During each test, we took photographs of the the bottom surface (Fig.\ \ref{fig:ProgressionOfDelamination}) and the lateral surface. From the images, we obtained a surface-averaged estimate of the radial fraction that had delaminated as a function of rotation angle (also shown in Fig.\ \ref{fig:ProgressionOfDelamination}). Note that only the initial delamination of a given part of a radial line was considered -- segments of the lines that had previously delaminated and re-adhered were still denoted as having delaminated. Corresponding measurements of the torque and normal force are shown in Fig. \ref{fig:TorqueAndNormalForce} (where compressive normal force values are positive). We used the images of the lateral surface to identify wrinkling instabilities and fracture initiation throughout the rotation, and after each test, we further examined the samples to determine the degree to which fractures had formed (see ESI). We identified and eliminated clear outlier tests based on these results (more detail in ESI).

Three sequential delamination phenomena emerge from our observations: \textit{(i)} First, peeling initiates at a site of imperfect adhesion between the sample and glass; \textit{(ii)} Next, the delamination forms a front that propagates along the circumference and radially inward in a stick-slip manner; \textit{(iii)} Finally, the radial progression of delamination is arrested when fractures begin to form on the lateral surface, leaving an adhered region in the center of the interface. These trends are found to be consistent in samples that are subjected to normal force; in absence of a normal force, delamination occurs more abruptly over the entire surface. These behaviors are discussed below in more detail:

\textit{(i) Initiation.} Across all the tests, peeling begins along the outer edge of the circular adhered surface at a site of imperfect adhesion between the sample and the glass plate, evident from the symmetry breaking in Fig. \ref{fig:ProgressionOfDelamination}. This pattern is in accordance with analytical models of the delamination of tubular adhesive joints \citep{shishesaz2019effects}, in which imperfections on the outer edge of the joint lead to higher stresses and increased likelihood of shear failure. Interestingly, our experiments show that the level of defect sensitivity in the initiation and propagation of delamination can be tuned by applying a normal force, as seen by the increasing repeatability among tests with increasing normal force in Fig. \ref{fig:TorqueAndNormalForce}.

\textit{(ii) Propagation.} Once delamination initiates, it propagates circumferentially, as seen by the shrinkage of the red-shaded regions in Fig. \ref{fig:ProgressionOfDelamination} with increasing rotation angle. The propagation occurs in a stick-slip manner, such that regions that have previously delaminated re-adhere and later delaminate again. For samples with no applied normal force, delamination propagates through the sample's entire radius from the onset of peeling (Fig. \ref{fig:ProgressionOfDelamination}a). For samples with an applied normal force, the stick-slip cycles start at the edge and spiral inward as they move around the circumference (Fig. \ref{fig:ProgressionOfDelamination}b,c). 
The propagation of delamination that we observe is in line with results from prior torsion studies: the inward-radial progression matches the findings of \citet{chateauminois2010friction}, and the circumferential stick-slip cycles are broadly consistent with the work of \citet{chaudhury2007studying}. This stick-slip process is visible as alternating jumps and plateaus in the delaminated radius percentage in Fig.\ \ref{fig:ProgressionOfDelamination} and corresponds to fluctuations in the otherwise-monotonic normal force and torque graphs (Fig. \ref{fig:TorqueAndNormalForce}; also see ESI for more detail). 

\textit{(iii) Arrest.} In all samples, we eventually observe the termination of delamination. The reason for the termination depends on the applied normal force. For samples with non-zero applied force, the delamination stops at some radial distance from the center, giving way to fracturing. For samples without applied compression, the entire sample delaminates by the end of the test and does not fracture. This behavior may be seen by contrasting the observations in Fig.\ \ref{fig:ProgressionOfDelamination}: in the absence of normal force, the sample fully delaminates by 92 degrees of rotation, while the plots for samples with an initial normal force reach a plateau prior to delaminating through the whole surface. 

Aside from the three observations detailed above, we identified experimental evidence of several other phenomena. In Fig.\ \ref{fig:TorqueAndNormalForce}, the normal compression is shown to increase as the rotation angle increases. This finding is consistent with the well-documented Poynting effect, which predicts that torsion elongates cylinders, so compression must be applied to preserve their height \citep{kanner2008extension, zurlo2020poynting}. At the very beginning of rotation, the normal force sometimes decreases before it begins to increase -- one possible explanation is that any initial normal force can cause the  cylinder to slip laterally at onset of loading. The slippage becomes more pronounced for larger samples, as shown in the ESI. In addition, we observed the emergence of wrinkles on the lateral surface of the cylinders once the sample was rotated beyond beyond a critical angle, in accordance with the theory proposed by \citet{ciarletta2014torsion}. While the fractures on the lateral surface have been discussed briefly above, more investigation is required to understand the particular fracturing patterns that emerge in these compression/torsion tests. More detail on the wrinkling and fracturing patterns we observed is provided in the ESI.

Based on the experimental data, delamination of a circular surface under combined axial compression and torsion is a process in which an initial imperfection in the adhesion between the PDMS sample and the glass plate creates a delamination front that propagates circumferentially and radially inward. The sample often re-adheres to the substrate after delamination, creating successive stick-slip waves. Fluctuations in the trends of normal force and torque as a function of the rotation angle correspond to snap-through events where delamination progresses rapidly. The delamination process is highly sensitive to imperfections, though application of a normal force is shown to reduce the level of imperfection sensitivity and make the process considerably more repeatable. Moreover, by resisting delamination, applied compression limits the range of delamination and eventually drives the sample to fracture rather than delaminate the innermost part of the surface. However, if no normal force is initially applied, the sample delaminates across the whole interface and does not fracture. The physical mechanisms behind these key observations are discussed through a theoretical lens below.

\section{Theoretical Framework} 
\label{Theoretical Framework} 

In this section we derive a theoretical model that can explain the experimental phenomena described in the previous section. A key assumption that we make, to simplify the kinematic description, is axial symmetry. Although our experimental observations exhibit delamination patterns that start locally at imperfections, and are thus not initially axially symmetric, we find that, with applied normal force, the delamination propagates circumferentially and stabilizes once it completes a whole circle. Hence, axial symmetry becomes a good approximation to capture the quasistatic process, allowing us to obtain insights on the observed phenomena.

\begin{figure}[h]
    \centering
    \includegraphics[width=0.45\textwidth]{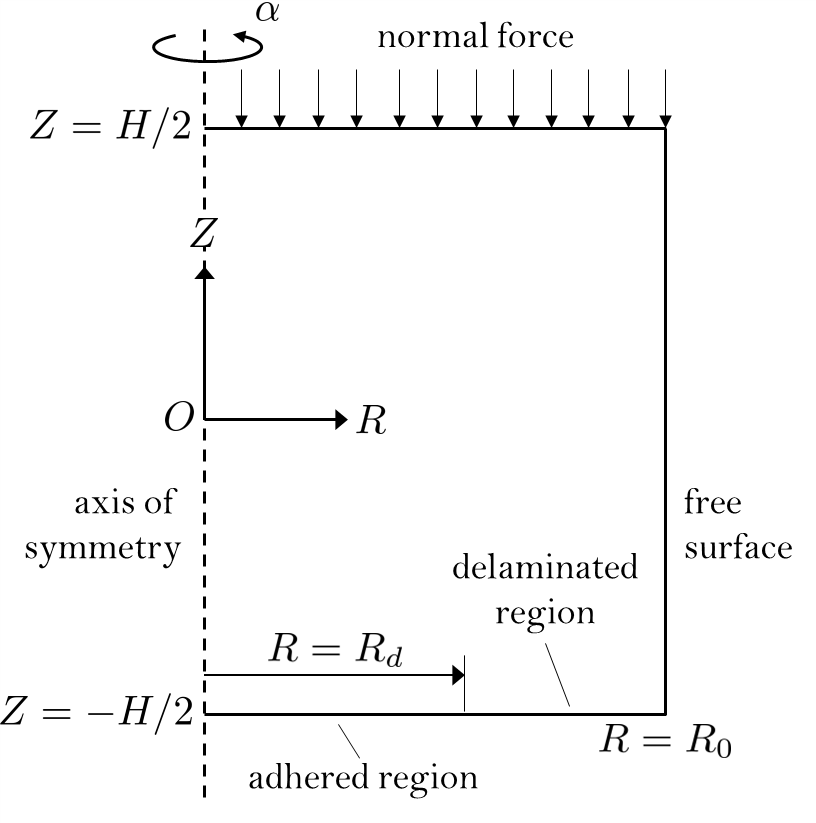} 
    \captionsetup{justification=centering}
    \caption{Illustration of the boundary value problem.}
    \label{fig:domain_illustration.png}
\end{figure}

\bigskip
\noindent \textbf{Problem setting and kinematic assumptions.} 
To describe the experimental system illustrated in Fig. \ref{fig:MainTextSchematic}, we consider an elastic cylinder made of a hyperelastic, isotropic, homogeneous, and incompressible material. Its undeformed axial length is denoted by $H$, and it has an undeformed radius of $R_{0}$. Material points are labeled using the cylindrical coordinate system, as $\boldsymbol{X}=\boldsymbol{X}(R,\Theta,Z)$ in the undeformed state, such that 
\begin{equation} 0\leq R\leq R_0, \qquad 0\leq\Theta< 2\pi, \qquad -H/2\leq Z \leq H/2. \end{equation} as illustrated in Fig. \ref{fig:domain_illustration.png}.
Upon deformation, a mapping function $\boldsymbol{x} = \boldsymbol{\chi}(\boldsymbol{X})$ assigns the material points to their location in the deformed cylindrical coordinates $\boldsymbol{x}=\boldsymbol{x}(r,\theta ,z)$.

The cylinder is initially pushed against a rigid substrate (at $Z=-H/2$), by application of a normal stress at the top surface ($Z=H/2$), which is then rotated quasistatically to increase the prescribed rotation angle $\alpha$. Delamination is permitted only at the bottom surface and, with the assumption of axial symmetry, is restricted to the circular region $R_d\leq R\leq R_0$, where the innermost radius of the delaminated region, $R_d=R_d(\alpha)$, can vary throughout the loading. Understanding the propagation of this delamination front and its dependence on model parameters is the central goal of this formulation. 

To simplify the mathematical derivation, a set of kinematic assumptions are made. First, we assume that horizontal planes remain horizontal throughout the deformation. Second, we impose axial symmetry such that the deformation field is independent of $\Theta$. As a result of these two assumptions, the vertical displacement $u_{z}$ of any point on a horizontal plane is independent of the radial and angular coordinates (i.e.  $u_{z}=u_z(Z)$). We restrict our attention to small initial vertical deformations imposed by the applied normal force, which results in a uniform vertical stretch ($\lambda_{z}={\partial z}/{\partial Z}$) that is held constant throughout the torsion process. Lastly, we assume that the cylinder preserves its cylindrical shape throughout the deformation process, namely that radial displacements are independent of the vertical coordinate, i.e. the radial displacement function becomes $u_r=u_r(R)$. This assumption neglects any barreling effects that may occur in the initial compression, essentially implying that the initial compression permits sliding along the substrate. Additionally, wrinkling of the free surface of the cylinder is neglected. In our experiments, such wrinkling may emerge at large rotations, as described in the previous section, but capturing the wrinkling process is beyond the scope of this theory. 
With these kinematic assumptions, we are left with displacement fields of the form
\begin{equation} 
r=R+u_{r}(R), \qquad \theta=\Theta+u_{\theta}(R,Z), \qquad z=Z+u_{z}(Z)
\end{equation} 
where the displacement functions are defined as the difference between the deformed (lowercase) and undeformed (uppercase) coordinates. The deformation gradient $\boldsymbol{F} = \partial \boldsymbol{\chi}/ \partial \boldsymbol{X}$ can thus be written as

\begin{equation}\boldsymbol{F} = \begin{bmatrix} \vspace{2mm}
\frac{\partial r}{\partial R} & 0 & 0 \\ \vspace{2mm}
 r \frac{\partial \theta}{\partial R} & \frac{r}{R} & r \frac{\partial \theta}{\partial Z}\\ 
0 & 0 &\lambda_z \end{bmatrix} \end{equation}

and incompressibility implies
\begin{equation} \label{FF}
\det \boldsymbol{F} =1 \quad \to \quad \frac{r}{R}\left(\frac{\partial r }{\partial R} \right) = \frac{1}{\lambda_{z}}.
\end{equation} 
Integrating the above formula, and eliminating radial translation of the cylinder (i.e. imposing $u_r(0)=0$), yields the relations\begin{equation} 
\frac{\partial r}{\partial R} = \frac{r}{R} = \frac{1}{\sqrt{\lambda_{z}}}
\end{equation} 
Substituting the above result into \eqref{FF}, we obtain the final form of the deformation gradient as 
\begin{equation}\label{FFF}
\boldsymbol{F} =\frac{1}{\sqrt{\lambda_{z}}} \begin{bmatrix} 
1 & 0 & 0 \\
R\left(\frac{\partial u_{\theta}}{\partial R}\right) & 1 & R\left(\frac{\partial u_{\theta}}{\partial vZ}\right)\\ 
 0 & 0 & \lambda_{z}^{3/2} 
\end{bmatrix} \end{equation}
where the only remaining unknown field variable is the angular displacement function $u_{\theta}$.

\bigskip
\noindent\textbf{Governing equations and boundary conditions.} The problem at hand is concerned with a non-conservative process, whereby torsional deformation induces delamination and frictional sliding. All of these energetic contributions must be considered to determine the propagation of delamination $R_d(\alpha)$. Nonetheless, if $R_d$ and $\alpha$ are separately prescribed, the deformation of the elastic body, $u_\theta(R,Z;R_d,\alpha)$ and the corresponding stored elastic energy $E_e=E_e(R_d,\alpha)$, can be derived for a known set of boundary conditions (note the notation used in defining $u_\theta$: the variables before the semicolon are spatial coordinates, while those after the semicolon are prescribed system parameters). After obtaining $E_e=E_e(R_d,\alpha)$, we will also consider the surface energy $E_d=E_d(R_d)$, i.e. the energy invested in delamination, and we will show in the next subsection that the work done by friction can be represented as an effective potential $E_f=E_f(R_d)$. By combining all of these energetic contributions, we will determine the location of the delamination front as the one that minimizes the total energy in the system. 

First, we employ the neo-Hookean model to describe the elastic strain energy density of the material as
\begin{equation}\label{psi_NH} 
\Psi = \frac{\mu}{2}(\text{tr}(\boldsymbol{F F}^{T} )-3) 
\end{equation} 
where $\mu$ is the shear modulus. Substituting \eqref{FFF} in \eqref{psi_NH} gives

\begin{equation} 
\Psi = \frac{\mu}{2}\left\{\frac{2}{\lambda_{z}}+ \frac{R^2}{\lambda_z}\left[\left(\frac{\partial u_{\theta}}{\partial R}\right)^2+\left(\frac{\partial u_{\theta}}{\partial Z}\right)^2\right]+\lambda_{z}^2-3\right\}
\end{equation} 
The total elastic energy of the system, $E_e$, is thus given by integration of the strain energy density over the entire volume of the cylinder: 
\begin{equation} \label{Ee} \begin{split}
E_e =\pi\mu \int\displaylimits_{-H/2}^{H/2} \int\displaylimits_{0}^{R_{0}}  \left\{\frac{2}{\lambda_{z}}+ \frac{R^2}{\lambda_z}\left[\left(\frac{\partial u_{\theta}}{\partial R}\right)^2+\left(\frac{\partial u_{\theta}}{\partial Z}\right)^2\right]+\lambda_{z}^2-3\right\} R {\rm d}R {\rm d}Z
 \end{split}\end{equation}

Next, the work $W$ done by the traction force on the top and bottom surfaces ($Z=-{H}/{2},~ {H}/{2}$), can be written as 
\begin{equation}\label{W}\qquad W =- 2\pi\int\displaylimits_{0}^{R_{0}} t_{\theta z} u_{\theta} R^{2} {\rm d}R \end{equation} 
where $t_{\theta z}$ is the generalized shear traction corresponding to displacements $u_\theta$, but is defined positive in the opposite direction. Here, as a first step and without loss of generality, we assume an arbitrary externally applied shear traction. Later we will specialize it to the different boundary regions (i.e. the regions of of applied torque and friction). 

An additional energy contribution $E_d$ comes from the creation of new surface area as the bottom surface of the cylinder delaminates -- this contribution can be written as
\begin{equation} \label{Ed}
E_{d}=\pi{\Gamma}(R_{0}^2-R_{d}^2)
\end{equation} where $\Gamma$ is the interface toughness per unit area.

An equilibrium solution minimizes the potential energy $U$ in the system, \begin{equation}\label{U}
U=E_e+E_d-W.\end{equation} 
For a prescribed rotation angle $\alpha$, the potential energy is a functional of $u_\theta$ and $R_d$, i.e. $U=U({u}_{\theta},R_d;\alpha)$. 

Using the tools of calculus of variations, we consider perturbations of the stationary function $u_{\theta}(R,Z)$ in the form $\tilde{u}_{\theta}=u_{\theta}+\epsilon \delta u_{\theta}$, where $\delta {u_{\theta}}(R,Z)$ is an arbitrary function that must vanish on the boundaries of the body that are subjected to a displacement constraint, and $\epsilon$ is a small constant. Next, we substitute $\tilde{u}_{\theta}$ into the potential energy and require that 

\begin{equation} 
\left.\frac{\partial {\tilde U}}{\partial \epsilon}\right\vert_{\epsilon \rightarrow 0} = 0, \qquad \text{and}\qquad \frac{\partial {U}}{\partial R_d} = 0
\label{dU} \end{equation} where we have used the notation ${\tilde U}=U(\tilde{u}_{\theta},R_d;\alpha)$.
By substituting \eqref{Ee} and \eqref{W}  and performing integration by parts, the first of the above two equations can be recast in the following form

\begin{equation}\label{integrals}\begin{split}
 \int\displaylimits_{-H/2}^{H/2} \int\displaylimits_{0}^{R_{0}} \left( 3 \frac{\partial u_{\theta}}{\partial R} + R\left(\frac{\partial^2 u_{\theta}}{\partial R^2}+\frac{\partial^2 u_{\theta}}{\partial Z^2}\right) \right)\delta u_\theta R^{2} {\rm d}R {\rm d}Z \qquad\qquad\qquad\\
- \int\displaylimits_{-H/2}^{H/2} \left. \left( \frac{\partial u_{\theta}}{\partial R} R^{3}\right)\delta u_{\theta}\right\vert_{R=R_{0}} {\rm d}Z 
\qquad\qquad\qquad\qquad\qquad\ \\ - \int\displaylimits_{0}^{R_{0}} \left. \left( \frac{\partial u_{\theta}}{\partial Z} R - \frac{\lambda_{z}}{\mu}t_{\theta z} \right)\delta u_{\theta}\right\vert_{Z=-\frac{H}{2},\frac{H}{2}} R^{2} {\rm d}R = 0 \qquad
\end{split}\end{equation}
which does not depend on the interfacial energy contribution $(E_d)$. For the above equality to hold for an arbitrary variation $\delta{u_{\theta}}$, each term must vanish separately. The first term gives rise to the governing equation 
\begin{equation}\label{GE}
\frac{\partial u_{\theta}}{\partial R} + \frac{R}{3}\left(\frac{\partial^2 u_{\theta}}{\partial R^2}+\frac{\partial^2 u_{\theta}}{\partial Z^2}\right)=0, \ \  R\in[0,R_0], \ \ Z\in\left[-\frac{H}{2},\frac{H}{2}\right] 
 \end{equation} 
The second integral implies that, on the free boundary of the cylinder, we have 
\begin{equation} \label{BC1} 
\frac{\partial u_{\theta}}{\partial R} = 0 \qquad\text{for}\quad R=R_0, \quad Z\in\left[-\frac{H}{2},\frac{H}{2}\right]
\end{equation} 

The last integral corresponds to the top and bottom surfaces of the cylinder. At $Z=H/2$, to enforce uniform torsion about the axis of symmetry, the displacement is prescribed as\begin{equation}\label{BC2}
 u_\theta=\alpha \qquad\text{for}\quad R\in[0,R_0], \quad Z=\frac{H}{2}
\end{equation} Hence, the variation must vanish ($\delta{u_{\theta}}=0$), and thus the corresponding integral term vanishes. At $Z=-H/2$, we split the integral to distinguish the different regions of the surface. In the adhered region, the displacement is prescribed: 
\begin{equation} \label{BC3}
 u_\theta=u_\theta^0(R) \qquad \text{for}\in[0,R_d], \quad Z=-\frac{H}{2}
\end{equation}
and thus the corresponding integral term vanishes. Note that initially $u_\theta^0=0$; however, if a `stick-slip' event occurs, bonding of the interface can be re-established, imposing a nonzero $u_\theta^0(R)$. This will be explained in more detail in describing the solution procedure.

In the delaminated region, sliding is permitted and can be balanced by a shear traction. The variation need not vanish in this region and thus, to ensure that the integral term vanishes, we enforce 
\begin{equation}\label{BC40}
t_{\theta z} = \mu\frac { R}{\lambda_{z}} \left( \frac{\partial u_{\theta}}{\partial Z}\right) \qquad\text{for}\quad R\in[R_d,R_0], \quad Z=-\frac{H}{2}
\end{equation}
The shear traction that emerges from the frictional sliding between the substrate and the cylinder acts as an external force on the system and needs to be prescribed to complete the boundary value problem. 

\bigskip
\noindent\textbf{Work of friction as an effective potential.} The work invested by external forces can be written as the sum of contributions from the applied torque, and the applied frictional force in the delaminated region, such that $W=W_t+W_f$, respectively. At the limit of quasistatic motion, where dynamic effects are neglected, we can simplify the representation of the kinetic friction by assuming that it is constant throughout the sliding, and proportional to the initially applied normal force $N$ through the constant $k$: $t_{\theta z}=kN$. Accordingly, \eqref{BC40} simplifies to a boundary condition on the slope of the displacement field, in the form
\begin{equation}\label{BC4}
\frac{\partial u_{\theta}}{\partial Z}=\frac{kN}{\mu R}\lambda_z\qquad\text{for}\quad R\in[R_d,R_0], \quad Z=-\frac{H}{2}
\end{equation}
Moreover, it is instructive to notice that  the work done by friction \eqref{W} can be alternatively written as 
\begin{equation}\label{Wf} Z=-\frac{H}{2}:\qquad W_f =-E_f=- 2\pi kN \int\displaylimits_{R_d}^{R_{0}} u_{\theta} R^{2} {\rm d}R \end{equation} 
which is path-independent\footnote{Note that here we restrict our attention to loading, without considering unloading.} and thus  $E_f=-W_f$ behaves as an effective potential, in analogy to the \textit{Rayleigh dissipation potential}. We can now also rewrite the potential energy in the form $U=E_e+E_d+E_f-W_t$. 


We would like to emphasize that although the present model is limited to quasistatic motion, kinetic friction must be accounted for here if unstable delamination ensues, and will allow us to capture the newly established equilibrium state that follows. This will be explained in more detail in describing the solution procedure. 

\vspace{2mm}
\noindent\textbf{Torque and normal force.} The above formulation is derived assuming a prescribed rotation angle -- $\alpha$, and deformed height -- $h=\lambda_z H$. Hence, the corresponding generalized forces, i.e. the torque -- $T$, and the normal force -- $N$ (defined positive in compression as shown in Fig. \ref{fig:MainTextSchematic}, where the compressive force $F$ is equivalent to the quantity $N$ discussed in this section), do not naturally emerge from the formulation. Nonetheless, a direct method to obtain them at a given state (i.e. with given $R_d$), is by considering perturbations of the elastic energy with respect to the corresponding generalized coordinate, which translates to the partial derivatives 
\begin{equation}
T = \frac{\partial E_e}{ \partial \alpha}, \qquad N =-\frac{\partial E_e}{ \partial h}= -\frac{1}{H} \frac{\partial E_e}{ \partial \lambda_{z}}
\end{equation} Though these derivatives can only be calculated a posteriori, they will be useful for comparison with the experimental observations. 

\vspace{2mm}
\noindent\textbf{Energy minimization.}
The above formulation consists of a single second order, partial differential equation \eqref{GE} for $u_\theta=u_\theta(R,Z)$, complemented by conditions on the different regions of the boundary \eqref{BC1}-\eqref{BC3} and \eqref{BC4}. 
This completes our boundary value problem, but for a prescribed location of the delamination front $R_d$ and for a given $\alpha$. To determine $R_d(\alpha)$, the location of the delamination front as rotation progresses, we must invoke the second requirement in \eqref{dU}. This requirement implies that the system should choose the location of the delamination front to minimize the total potential energy of the system. Though the dependence on $R_d$ is not explicit in $U$ and thus the differentiation cannot be done analytically, identifying the location of the front can be achieved numerically by considering all different candidate values of $R_d$ and finding the one that corresponds to the minimal potential energy. This can alternatively be written in the form
\begin{equation} R_{d}={\rm arg}~ \underset{\hat{R}_{d}}{\rm min} ~U(\hat{R}_{d}). \label{argmin_dimensional} \end{equation}

\noindent\textbf{Solution procedure and snap-through events.}
The delamination process is not gradual. The build-up of elastic energy in the cylinder is necessary to trigger delamination, which is then abrupt, leading to finite sliding at the interface, before arriving at a new equilibrium state for which kinetic friction is balanced by the elastic forces. Then, once at rest, the adhesive bond is re-established, akin to a static friction, and loading can proceed without immediately inducing additional delamination. It is instructive to notice that this sequence of events is analogous to a one-dimensional stick-slip system, with the applied displacement in the one-dimensional system representing the prescribed angular rotation of the top surface; the deformation of the spring in the one-dimensional system representing the torsional deformation of the cylinder; and the resulting uniaxial motion in the one-dimensional system representing the motion of material particles distributed throughout the surface, along circular trajectories.

\begin{figure*}[ht!]
    \centering
    \includegraphics[width=0.75\textwidth]{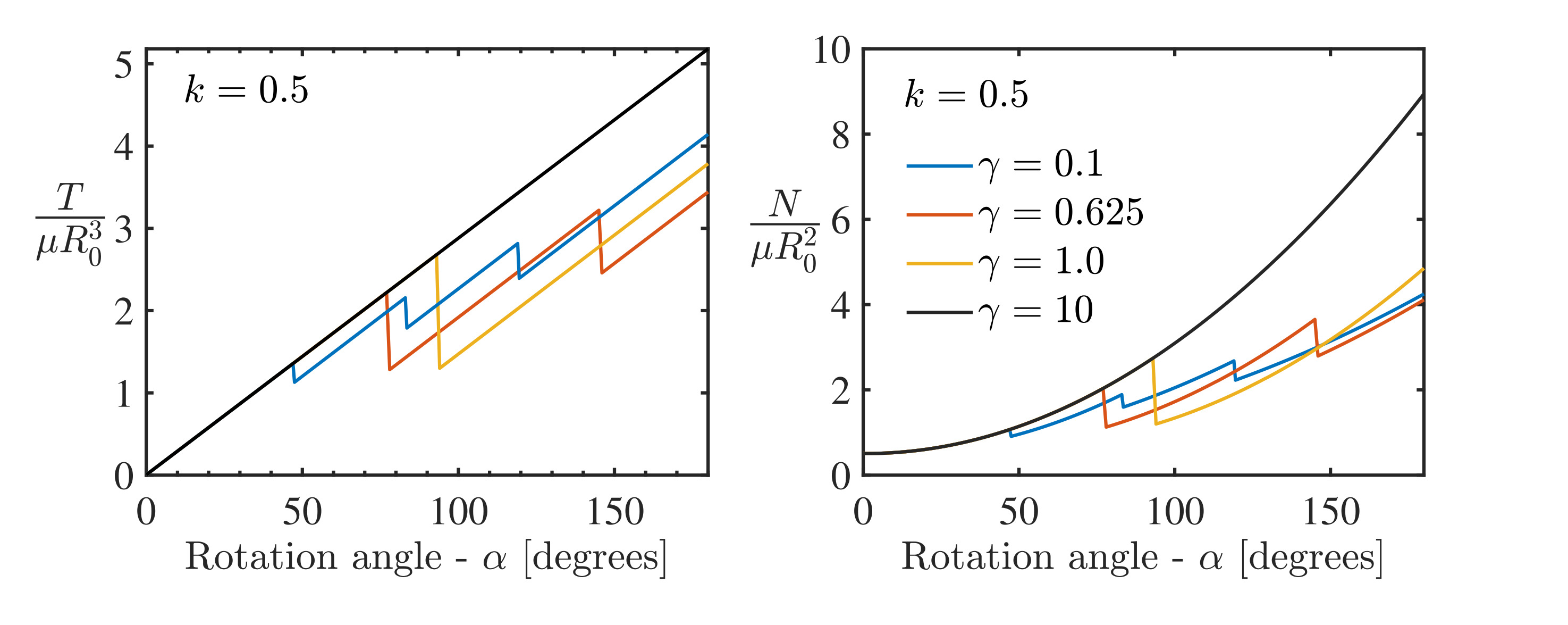} 
    \captionsetup{justification=centering}
    \caption{The evolution of dimensionless torque (left) and corresponding normal force (right) with respect to rotation angle, for various values of $\gamma$. Here we use $k=0.5$, $R_0/H=1$, and the initially applied dimensionless normal force is set to $0.5$. }
    \label{fig:g} 
\end{figure*}
\begin{figure*}[ht!]
    \centering
    \includegraphics[width=0.75\textwidth]{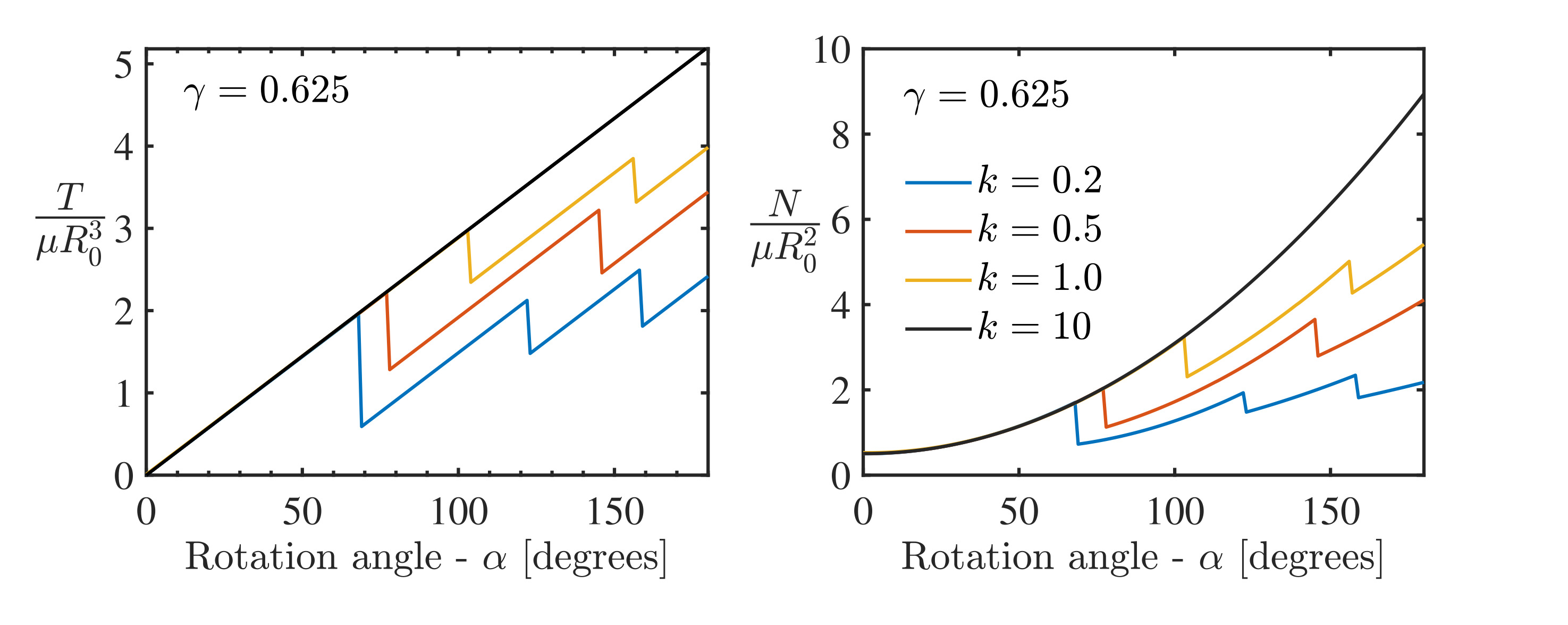} 
    \captionsetup{justification=centering}
    \caption{The evolution of dimensionless torque (left) and corresponding normal force (right) with respect to rotation angle, for various values of $k$. Here we use $\gamma=0.625$, $R_0/H=1$, and the initially applied dimensionless normal force is set to $0.5$.}
    \label{fig:k} 
\end{figure*}

In each of the steps of the deformation and delamination process described above (i.e. as $\alpha$ increases), the boundary conditions need to be adjusted accordingly in the solution procedure:

\smallskip
\noindent{\textit{Step I - Initial deformation:}} In this initial stage of the process, no delamination has occurred. Accordingly, the boundary value problem is solved with $R_d=R_0$ and $u_\theta^0=0$. 
Nonetheless, at every increment of applied rotation $\alpha$, solutions with $R_d\leq R_0$ are examined, employing the shear traction boundary condition \eqref{BC4}. Then, the minimum energy requirement \eqref{argmin_dimensional} is used to determine the location of the propagation front. If $R_d=R_0$, the solution procedure remains in Step I; otherwise, it transitions to Step II.

\noindent{\textit{Step II - Snap-through:}} Once $R_d=R_0$ no longer provides an energetically favorable solution, a delamination front will propagate, also leading to a drop in the applied torque. Considering a quasistatic process, our model captures the new equilibrium state that the system will choose, which results from the minimum energy requirement \eqref{argmin_dimensional}, as calculated in the previous step. Once the propagation stops, the adhesive bond is re-established and the displacement at the bottom surface is constrained to a new location, namely $u_\theta^0$ is re-assigned to $u_\theta^0(R)=u_\theta(R,-H/2)$.

\smallskip
\noindent{\textit{Step III - Secondary deformation:}} In this step torsional deformation proceeds with no additional delamination. Hence, the boundary value problem is solved with $R_d=R_0$ and $u_\theta^0(R)$ from Step II. Similar to Step I, in this stage of the process, no additional delamination occurs. Nonetheless, at every increment of applied rotation $\alpha$ we examine solutions with $R_d \leq R_0$ and employ the shear traction boundary condition \eqref{BC4}. The minimum energy requirement \eqref{argmin_dimensional} is again used to determine the location of the propagation front. If $R_d=R_0$ then the solution procedure remains in Step III, otherwise it transitions back to Step II.

All of the numerical derivations are conducted in Matlab. The governing equation in \eqref{GE} is integrated using a finite difference scheme.


\section{Results and Sensitivity Analysis} 

\label{Results and Sensetivity Analysis} 

To study the delamination process, we employ the solution procedure described in the previous section. The constitutive sensitivities are investigated by varying the dimensionless interface toughness $\gamma=\Gamma/\mu R_0$ and the friction coefficient $k$, while holding the dimensions of our sample constant with $R_0/H=1$. Motivated by the experimental observations, we restrict our attention to a normal compression value of $N/\mu R_0^2=0.5$. For comparison, the experimental values are $R_0/H=1.2$ and $N/\mu R_0^2=0.43$, as calculated with $\mu=E/3=8[kPa]$ and with $R_0=120$ mm.

Though we do not attempt to fit and directly compare between the experiments and the model, which is limited to cylindrically symmetric deformations and thus cannot capture the transient propagation of delamination along the circumference, we observe similar trends. We find that a good qualitative agreement between the experiments in Fig. \ref{fig:TorqueAndNormalForce} and the model in Figs. \ref{fig:g} and \ref{fig:k} is achieved for $k=0.5$ and $\gamma=0.625$ (red curves). This implies $\Gamma\sim 60~J/m^2$, which agrees with the reported surface toughness in the literature \citep{wahdat2022pressurized}. Other curves correspond to varying levels of surface toughness (Fig. \ref{fig:g}) and friction coefficient (Fig. \ref{fig:k}).

Both the dimensionless torque and the dimensionless normal force initially increase monotonically with rotation $\alpha$. The linear torque response is explained by the linearity of the neo-Hookean model under shear deformation, and the increase in the normal force is explained by the Poynting effect in large deformations of soft solids \citep{kanner2008extension, zurlo2020poynting}. The sharp drops observed in the curves indicate delamination events, which in the present framework manifest as first-order transitions and thus exhibit a finite drop in the applied load, until a new equilibrium state is achieved by the resistance of the frictional force. 
Such `snap-through' events are observed in all the curves except for $\gamma=10$, for which the energetic cost of delamination is too high and no delamination is observed within this range of $\alpha$. 
For lower values of $\gamma$ and for the range of considered friction coefficients, $k$, we observe that weaker interfaces (analogous to the cases with low normal force in the experiments) exhibit higher number of stick-slip events. 
 
\begin{figure}[h]
    \includegraphics[width=0.5\textwidth]{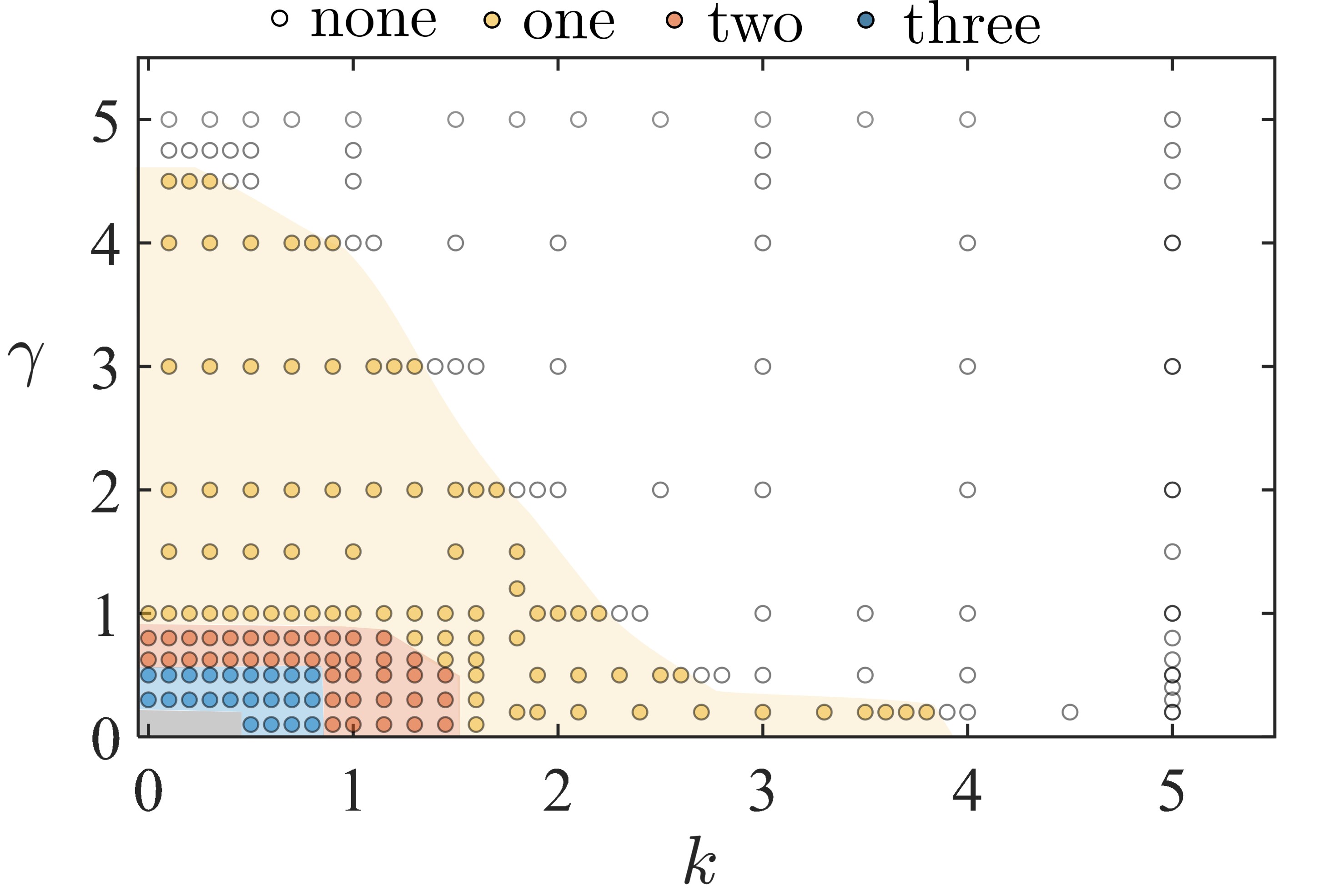} 
    \captionsetup{justification=centering}
    \caption{Phase portrait showing the number of stick-slip events that occur for rotation of $\alpha=180$ degrees, as a function of the dimensionless material properties $(k,\gamma)$. Different shaded regions correspond to different numbers of delamination events, as identified from theoretical predictions, indicated via circular markers. The gray region corresponds to situations in which the number of snap-through events is greater than $4$.}
    \label{phase} 
\end{figure}

 The influence of material parameters in determining the occurrence of snap-through events is further examined using the phase portrait in Fig. \ref{phase}. Each point on the phase portrait, associated with a pair $(k,\gamma)$, corresponds to a theoretical prediction for a system that has reached $\alpha=180$ degrees. The number of delamination events within this range is then recorded using the corresponding color. A clear trend emerges, where the number of delamination events increases for weaker interfaces (i.e. smaller $\gamma$ and $k$) until a high number of consecutive stick-slip cycles take place, thus arriving at a limit where frictional sliding and stick-slip become indistinguishable.

\section{Conclusion}
\label{section:Conclusion}

Soft adhesive solids are becoming ubiquitous due to their advantages over traditional joining techniques. This renders the understanding of their mechanical behavior and failure under these settings vital. Here, relevant physical phenomena to their mechanical behavior, failure and self-healing under combined compression-torsion loading are experimentally investigated and reported. A minimal theoretical framework is provided as an attempt to have a basic understanding of the observed phenomena, and it is shown to be in qualitative agreement with the experimental observations. There is a breadth of
opportunities to extend the theoretical model to include fracture in the bulk, examine the response for different constitutive models that incorporate strain stiffening, viscoelasticity, and compressibility, to investigate the change in adhesive strength from one delamination to the next, and to expand the framework to capture the role of inertia or non-rigid substrates. In addition, the experiments could be broadene to test a greater range of radius-to-height aspect ratios and different PDMS base:cross-linker ratios, to determine how changes in the geometric and material parameters may affect the properties of the adhesive bond and the delamination phenomena discussed above.

\appendix

\section*{Acknowledgements} The authors acknowledge the support from the National Science Foundation under award number CMMI-1942016. T.K.V.         acknowledges the Barry Goldwater Scholarship and the support of the MIT Undergraduate Research Opportunities Program.
\section*{References}
\bibliography{rsc} 

\begin{thebibliography}{40}
\providecommand{\natexlab}[1]{#1}
\providecommand{\url}[1]{\texttt{#1}}
\expandafter\ifx\csname urlstyle\endcsname\relax
  \providecommand{\doi}[1]{doi: #1}\else
  \providecommand{\doi}{doi: \begingroup \urlstyle{rm}\Url}\fi

\bibitem[Alfano et~al.(2018)Alfano, Morano, Moroni, Musiari, Spennacchio, and
  Di~Lonardo]{alfano2018fracture}
Marco Alfano, Chiara Morano, Fabrizio Moroni, Francesco Musiari,
  Giuseppe~Danilo Spennacchio, and Donato Di~Lonardo.
\newblock Fracture toughness of structural adhesives for the automotive
  industry.
\newblock \emph{Procedia Structural Integrity}, 8:\penalty0 561--565, 2018.

\bibitem[Kinloch(1997)]{kinloch1997adhesives}
AJ~Kinloch.
\newblock Adhesives in engineering.
\newblock \emph{Proceedings of the Institution of Mechanical Engineers, Part G:
  Journal of Aerospace Engineering}, 211\penalty0 (5):\penalty0 307--335, 1997.

\bibitem[Davies et~al.(2006)Davies, Hitchings, and
  Ankersen]{davies2006predicting}
GAO Davies, D~Hitchings, and J~Ankersen.
\newblock Predicting delamination and debonding in modern aerospace composite
  structures.
\newblock \emph{Composites Science and Technology}, 66\penalty0 (6):\penalty0
  846--854, 2006.

\bibitem[Barzegar and Mokhtari(2019)]{barzegar2019numerical}
Mohsen Barzegar and Majid Mokhtari.
\newblock Numerical study of geostationary orbit thermal cycle effects of a
  tubular adhesive joint: Dynamic behavior.
\newblock \emph{The Journal of Adhesion}, 2019.

\bibitem[Mays and Hutchinson(1992)]{mays1992adhesives}
Geoffrey~C Mays and Allan~R Hutchinson.
\newblock \emph{Adhesives in civil engineering}, volume~32.
\newblock Cambridge University Press Cambridge, UK:, 1992.

\bibitem[Buyukozturk and Hearing(1998)]{buyukozturk1998failure}
Oral Buyukozturk and Brian Hearing.
\newblock Failure behavior of precracked concrete beams retrofitted with frp.
\newblock \emph{Journal of composites for construction}, 2\penalty0
  (3):\penalty0 138--144, 1998.

\bibitem[Rahimi and Hutchinson(2001)]{rahimi2001concrete}
Hamid Rahimi and Allan Hutchinson.
\newblock Concrete beams strengthened with externally bonded frp plates.
\newblock \emph{Journal of composites for construction}, 5\penalty0
  (1):\penalty0 44--56, 2001.

\bibitem[Buyukozturk et~al.(2004)Buyukozturk, Gunes, and
  Karaca]{buyukozturk2004progress}
Oral Buyukozturk, Oguz Gunes, and Erdem Karaca.
\newblock Progress on understanding debonding problems in reinforced concrete
  and steel members strengthened using frp composites.
\newblock \emph{Construction and Building Materials}, 18\penalty0 (1):\penalty0
  9--19, 2004.

\bibitem[Hartshorn(2012)]{hartshorn2012structural}
Stephen~Richard Hartshorn.
\newblock \emph{Structural adhesives: chemistry and technology}.
\newblock Springer Science \& Business Media, 2012.

\bibitem[Li et~al.(2016)Li, Krahn, and Menon]{li2016bioinspired}
Yasong Li, Jeffrey Krahn, and Carlo Menon.
\newblock Bioinspired dry adhesive materials and their application in robotics:
  a review.
\newblock \emph{Journal of Bionic Engineering}, 13\penalty0 (2):\penalty0
  181--199, 2016.

\bibitem[Jiang et~al.(2017)Jiang, Hawkes, Fuller, Estrada, Suresh, Abcouwer,
  Han, Wang, Ploch, Parness, et~al.]{jiang2017robotic}
Hao Jiang, Elliot~W Hawkes, Christine Fuller, Matthew~A Estrada, Srinivasan~A
  Suresh, Neil Abcouwer, Amy~K Han, Shiquan Wang, Christopher~J Ploch, Aaron
  Parness, et~al.
\newblock A robotic device using gecko-inspired adhesives can grasp and
  manipulate large objects in microgravity.
\newblock \emph{Science Robotics}, 2\penalty0 (7), 2017.

\bibitem[Bowditch and Shaw(1996)]{bowditch1996adhesive}
MR~Bowditch and SJ~Shaw.
\newblock Adhesive bonding for high performance materials.
\newblock \emph{Advanced Performance Materials}, 3\penalty0 (3-4):\penalty0
  325--342, 1996.

\bibitem[Shang et~al.(2019)Shang, Marques, Machado, Carbas, Jiang, and
  da~Silva]{shang2019strategy}
X~Shang, EAS Marques, JJM Machado, RJC Carbas, D~Jiang, and LFM da~Silva.
\newblock A strategy to reduce delamination of adhesive joints with composite
  substrates.
\newblock \emph{Proceedings of the Institution of Mechanical Engineers, Part L:
  Journal of Materials: Design and Applications}, 233\penalty0 (3):\penalty0
  521--530, 2019.

\bibitem[Banea and da~Silva(2010)]{banea2010effect}
MD~Banea and Lucas~FM da~Silva.
\newblock The effect of temperature on the mechanical properties of adhesives
  for the automotive industry.
\newblock \emph{Proceedings of the Institution of Mechanical Engineers, Part L:
  Journal of Materials: Design and Applications}, 224\penalty0 (2):\penalty0
  51--62, 2010.

\bibitem[Cohen et~al.(2018)Cohen, Chan, and Mahadevan]{cohen2018competing}
Tal Cohen, Chon~U Chan, and L~Mahadevan.
\newblock Competing failure modes in finite adhesive pads.
\newblock \emph{Soft matter}, 14\penalty0 (10):\penalty0 1771--1779, 2018.

\bibitem[Collino et~al.(2014)Collino, Philips, Rossol, McMeeking, and
  Begley]{collino2014detachment}
Rachel~R Collino, Noah~R Philips, Michael~N Rossol, Robert~M McMeeking, and
  Matthew~R Begley.
\newblock Detachment of compliant films adhered to stiff substrates via van der
  waals interactions: role of frictional sliding during peeling.
\newblock \emph{Journal of The Royal Society Interface}, 11\penalty0
  (97):\penalty0 20140453, 2014.

\bibitem[Ringoot et~al.(2021)Ringoot, Roch, Molinari, Massart, and
  Cohen]{ringoot2021stick}
Evelyne Ringoot, Thibault Roch, Jean-Fran{\c{c}}ois Molinari, Thierry~J
  Massart, and Tal Cohen.
\newblock Stick--slip phenomena and schallamach waves captured using reversible
  cohesive elements.
\newblock \emph{Journal of the Mechanics and Physics of Solids}, 155:\penalty0
  104528, 2021.

\bibitem[Shull et~al.(1998)Shull, Ahn, Chen, Flanigan, and
  Crosby]{shull1998axisymmetric}
Kenneth~R Shull, Dongchan Ahn, Wan-Lin Chen, Cynthia~M Flanigan, and Alfred~J
  Crosby.
\newblock Axisymmetric adhesion tests of soft materials.
\newblock \emph{Macromolecular Chemistry and Physics}, 199\penalty0
  (4):\penalty0 489--511, 1998.

\bibitem[Oh(2007)]{oh2007strength}
Je~Hoon Oh.
\newblock Strength prediction of tubular composite adhesive joints under
  torsion.
\newblock \emph{Composites science and technology}, 67\penalty0 (7-8):\penalty0
  1340--1347, 2007.

\bibitem[Liu et~al.(2018)Liu, He, Hamon, Fan, Haghi-Ashtiani, Reiss, and
  Bai]{liu2018comparison}
Yu~Liu, Delong He, Ann-Lenaig Hamon, Benhui Fan, Paul Haghi-Ashtiani, Thomas
  Reiss, and Jinbo Bai.
\newblock Comparison of different surface treatments of carbon fibers used as
  reinforcements in epoxy composites: Interfacial strength measurements by
  in-situ scanning electron microscope tensile tests.
\newblock \emph{Composites Science and Technology}, 167:\penalty0 331--338,
  2018.

\bibitem[Khashaba et~al.(2020)Khashaba, Othman, and
  Najjar]{khashaba2020development}
UA~Khashaba, Ramzi Othman, and Ismael~MR Najjar.
\newblock Development and characterization of structural adhesives for
  aerospace industry with alumina nanoparticles under shear and
  thermo-mechanical impact loads.
\newblock \emph{Proceedings of the Institution of Mechanical Engineers, Part G:
  Journal of Aerospace Engineering}, 234\penalty0 (2):\penalty0 490--507, 2020.

\bibitem[Razavi et~al.(2017)Razavi, Ayatollahi, Esmaeili, and
  Da~Silva]{razavi2017mixed}
SMJ Razavi, MR~Ayatollahi, E~Esmaeili, and LFM Da~Silva.
\newblock Mixed-mode fracture response of metallic fiber-reinforced epoxy
  adhesive.
\newblock \emph{European Journal of Mechanics-A/Solids}, 65:\penalty0 349--359,
  2017.

\bibitem[Bartlett et~al.(2012)Bartlett, Croll, King, Paret, Irschick, and
  Crosby]{bartlett2012looking}
Michael~D Bartlett, Andrew~B Croll, Daniel~R King, Beth~M Paret, Duncan~J
  Irschick, and Alfred~J Crosby.
\newblock Looking beyond fibrillar features to scale gecko-like adhesion.
\newblock \emph{Advanced Materials}, 24\penalty0 (8):\penalty0 1078--1083,
  2012.

\bibitem[Chateauminois et~al.(2010)Chateauminois, Fretigny, and
  Olanier]{chateauminois2010friction}
Antoine Chateauminois, Christian Fretigny, and Ludovic Olanier.
\newblock Friction and shear fracture of an adhesive contact under torsion.
\newblock \emph{Physical Review E}, 81\penalty0 (2):\penalty0 026106, 2010.

\bibitem[Hetenyi and McDonald~Jr(1958)]{hetenyi1958contact}
M~Hetenyi and PH~McDonald~Jr.
\newblock Contact stresses under combined pressure and twist.
\newblock 1958.

\bibitem[Bryant and Dukes(1965)]{bryant1965measurement}
RW~Bryant and WA~Dukes.
\newblock The measurement of the shear strength of adhesive joints in torsion.
\newblock \emph{British Journal of Applied Physics}, 16\penalty0 (1):\penalty0
  101, 1965.

\bibitem[Shishesaz and Tehrani(2019)]{shishesaz2019effects}
Mohammad Shishesaz and Siavash Tehrani.
\newblock The effects of circumferential voids or debonds on stress
  distribution in tubular adhesive joints under torsion.
\newblock \emph{The Journal of Adhesion}, 2019.

\bibitem[Chaudhury and Chung(2007)]{chaudhury2007studying}
Manoj~K Chaudhury and Jun~Young Chung.
\newblock Studying friction and shear fracture in thin confined films using a
  rotational shear apparatus.
\newblock \emph{Langmuir}, 23\penalty0 (15):\penalty0 8061--8066, 2007.

\bibitem[Henzel et~al.(2022)Henzel, Nijjer, Chockalingam, Wahdat, Crosby, Yan,
  and Cohen]{interfacial}
Thomas Henzel, Japinder Nijjer, S~Chockalingam, Hares Wahdat, Alfred~J Crosby,
  Jing Yan, and Tal Cohen.
\newblock {Interfacial cavitation}.
\newblock \emph{PNAS Nexus}, 10 2022.
\newblock ISSN 2752-6542.
\newblock \doi{10.1093/pnasnexus/pgac217}.
\newblock URL \url{https://doi.org/10.1093/pnasnexus/pgac217}.
\newblock pgac217.

\bibitem[Wahdat et~al.(2022)Wahdat, Zhang, Chan, and
  Crosby]{wahdat2022pressurized}
Hares Wahdat, Cathy Zhang, Nicky Chan, and Alfred~J Crosby.
\newblock Pressurized interfacial failure of soft adhesives.
\newblock \emph{Soft Matter}, 18\penalty0 (4):\penalty0 755--761, 2022.

\bibitem[P{\'e}rez-R{\`a}fols and Nicola(2021)]{perez2021incipient}
Francesc P{\'e}rez-R{\`a}fols and Lucia Nicola.
\newblock Incipient sliding of adhesive contacts.
\newblock \emph{Friction}, pages 1--14, 2021.

\bibitem[Barquins et~al.(1976)Barquins, Courtel, and
  Maugis]{barquins1976friction}
M~Barquins, R~Courtel, and D~Maugis.
\newblock Friction on stretched rubber.
\newblock \emph{Wear}, 38\penalty0 (2):\penalty0 385--389, 1976.

\bibitem[Schallamach(1971)]{schallamach1971does}
A~Schallamach.
\newblock How does rubber slide?
\newblock \emph{Wear}, 17\penalty0 (4):\penalty0 301--312, 1971.

\bibitem[Kanner and Horgan(2008)]{kanner2008extension}
Landon~M Kanner and Cornelius~O Horgan.
\newblock On extension and torsion of strain-stiffening rubber-like elastic
  circular cylinders.
\newblock \emph{Journal of Elasticity}, 93\penalty0 (1):\penalty0 39--61, 2008.

\bibitem[Zurlo et~al.(2020)Zurlo, Blackwell, Colgan, and
  Destrade]{zurlo2020poynting}
Giuseppe Zurlo, James Blackwell, Niall Colgan, and Michel Destrade.
\newblock The poynting effect.
\newblock \emph{American Journal of Physics}, 88\penalty0 (12):\penalty0
  1036--1040, 2020.

\bibitem[Ciarletta and Destrade(2014)]{ciarletta2014torsion}
Pasquale Ciarletta and Michel Destrade.
\newblock Torsion instability of soft solid cylinders.
\newblock \emph{The IMA Journal of Applied Mathematics}, 79\penalty0
  (5):\penalty0 804--819, 2014.

\bibitem[Raayai-Ardakani et~al.(2019)Raayai-Ardakani, Chen, Earl, and
  Cohen]{raayai2019volume}
Shabnam Raayai-Ardakani, Zhantao Chen, Darla~Rachelle Earl, and Tal Cohen.
\newblock Volume-controlled cavity expansion for probing of local elastic
  properties in soft materials.
\newblock \emph{Soft matter}, 15\penalty0 (3):\penalty0 381--392, 2019.

\bibitem[Chockalingam et~al.(2021)Chockalingam, Roth, Henzel, and
  Cohen]{chockalingam2021probing}
S~Chockalingam, Christine Roth, Thomas Henzel, and Tal Cohen.
\newblock Probing local nonlinear viscoelastic properties in soft materials.
\newblock \emph{Journal of the Mechanics and Physics of Solids}, 146:\penalty0
  104172, 2021.

\bibitem[Faley et~al.(2015)Faley, Baer, Larsen, and Bellan]{faley2015robust}
Shannon~L Faley, Bradly~B Baer, Taylor~SH Larsen, and Leon~M Bellan.
\newblock Robust fluidic connections to freestanding microfluidic hydrogels.
\newblock \emph{Biomicrofluidics}, 9\penalty0 (3):\penalty0 036501, 2015.

\bibitem[Gent and Hua(2004)]{gent2004torsional}
AN~Gent and K-C Hua.
\newblock Torsional instability of stretched rubber cylinders.
\newblock \emph{International Journal of Non-Linear Mechanics}, 39\penalty0
  (3):\penalty0 483--489, 2004.

\end{thebibliography}


\newpage
\setcounter{page}{1}
\fancyfoot{}
\fancyfoot[LO,RE]{}
\fancyfoot[CO]{}
\fancyfoot[CE]{}
\fancyfoot[RO]{\footnotesize{\sffamily{~\textbar  \hspace{2pt}~\thepage}}}
\fancyfoot[LE]{\footnotesize{\sffamily{\thepage\hspace{2pt} ~\textbar}}}
\fancyhead{}
\renewcommand{\headrulewidth}{0pt} 
\renewcommand{\footrulewidth}{0pt}
\setlength{\arrayrulewidth}{1pt}
\setlength{\columnsep}{1mm}
\setlength\bibsep{1pt}

\makeatother

\onecolumn

{
\baselineskip = 16pt

\begin{center}
\textbf{\Large Electronic Supplementary Information for the Manuscript:}

\textbf{\Large ``Torsional Instabilities in the Delamination of Soft Adhesives"}

Tara K. Venkatadri, Thomas Henzel, and Tal Cohen
\end{center}
}


\setcounter{equation}{0}
\setcounter{figure}{0}
\setcounter{table}{0}
\setcounter{section}{0}
\setcounter{subsection}{0}
\makeatletter
\renewcommand{\theequation}{S\arabic{equation}}
\renewcommand{\thefigure}{SI.\arabic{section}.\arabic{figure}}
\renewcommand{\thesection}{ESI.\arabic{section}}
\renewcommand{\thesubsection}{ESI.\arabic{subsection}}

\setcounter{section}{0}
\makeatletter
\renewcommand{\thesection}{ESI.\arabic{section}}
\renewcommand{\thesubsection}{ESI.\arabic{subsection}}

\section{Delamination Experiment with Larger Sample}
\label{appendix:LargerSample}
\setcounter{figure}{0} 
The analysis in the main paper was restricted to small samples with height $\sim 1 $ cm and diameter $\sim 2.4$ cm. To examine the influence of the sample's geometry on the delamination process, we conducted a similar experiment on a larger sample, with a diameter of $7.62$ cm ($3$ inches) and a height of $0.95$ cm ($0.375$ inches). As with the experiments on smaller samples, we used PDMS with a base to cross-linker ratio of 40:1. We used the same experimental setup described in \ref{appendix:ExperimentalSetupAndProtocol}, although the size of this sample required the use of a wider acrylic disk mounted to the Instron. Given the size of the sample, $16$ radial lines were drawn instead of $8$ (which was the case for the smaller samples), to track delamination more precisely.

\begin{figure*}
    \centering
    \includegraphics[width=0.8\textwidth]{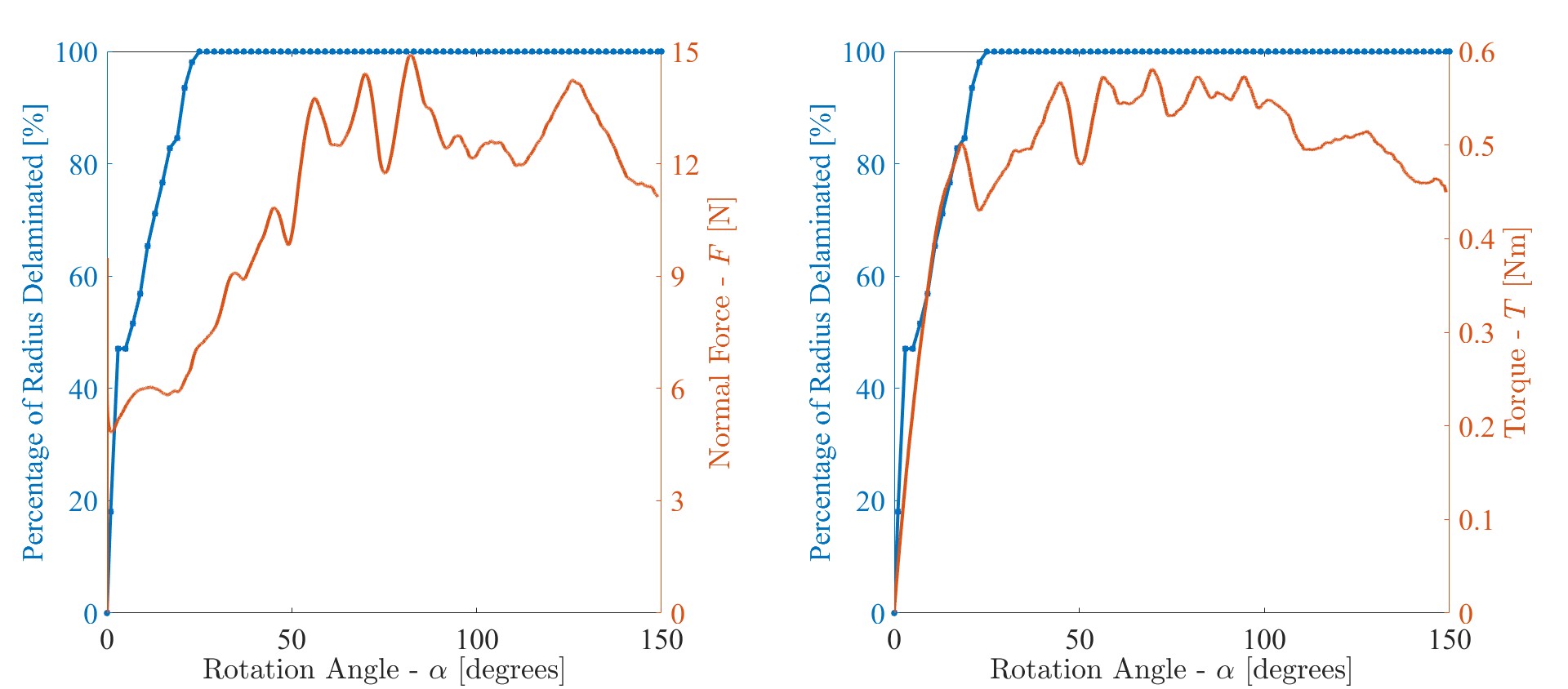} 
    \captionsetup{justification=centering}
    \caption{Delaminated radius as a function of the rotation angle, overlaid on graphs of normal force (left) and torque (right).  The sample delaminated rapidly, reaching 100\% delamination by a rotation angle of 25 degrees. It did not exhibit the stick-slip cycles seen for the smaller samples. 
    }
    \label{fig:LargeSampleAppendixGraphs}
\end{figure*}

One key parameter that required adjustment for the larger test was the normal force. Since this sample had a contact area roughly $10$ times that of the smaller samples, we applied 5 N of compression, $10$ times the initial normal force, to provide the same axial stress as the 0.5 N tests in the main text. A time-lapse video of the delamination process (one version with the delamination of the radial lines marked and color-coded, and another version without markings) is available at the following \href{https://drive.google.com/drive/folders/1XjKZ8HKTsY8BhSIHX8Wy9IqrMPHnHHpo?usp=sharing}{link}. Based on the results of this test, we determined how the contact area affects the three delamination processes described in Section \ref{section:ExperimentalObservations}:



Fig. \ref{fig:LargeSampleAppendixGraphs} provides additional insight into the delamination process. As a result of the rapid delamination, the stick-slip patterns are not visible as breaks in slope in the plot of delaminated radius percentage as a function of rotation angle, and instead there is one continuous `slip' regime from the start of the rotation until the sample has fully delaminated. Prior to full delamination, the torque and normal force follow trends broadly similar to Fig. 3 in the main text. After full delamination occurs, they fluctuate about a mean value, consistent with the stick-slip and frictional sliding behavior described below.

\underline{\textit{No arrest}:} 
Unlike the smaller samples with 0.5 N of applied normal force, to which this test was analogous, the larger sample did not stop delaminating and give way to fracturing at any point. Fractures primarily developed along localized imperfections, rather than forming around the circumference of the sample as with the smaller samples. Instead of transitioning to a fracturing-dominated regime, the bottom of the sample continued to rotate via frictional sliding. The outer edges of the radial lines moved as quickly as the Instron load cell, entering a state of nearly-rigid body motion. The inner sections of the lines (which experienced smaller strains, akin to what was observed in the smaller samples) moved in a somewhat stick-slip manner and did not move as a rigid body with the load cell, so the initially-straight radial lines became curved.

\underline{\textit{Imperfection sensitivity}:} The most prominent difference between this sample and the smaller samples, which drove many of the observations described above, was the increased sensitivity of the larger sample to imperfections in the experimental setup, namely the fact that the tabletop was not completely level. This imperfection caused the radial lines on the top half of the \href{https://drive.google.com/drive/folders/1XjKZ8HKTsY8BhSIHX8Wy9IqrMPHnHHpo?usp=sharing}{video} to delaminate more freely and experience frictional sliding, while the lines on the bottom half were more likely to undergo stick-slip motion. The sample's size provides a large area over which a small tilt in the plates (such that they are not perfectly parallel), or an imperfect thickness distribution of the sample itself, could act.
For larger samples with radius $R_0$, height $H$, and applied normal force $F$, a slight tilt angle $\beta$ in the adhered surface induces a change in the height of the sample’s edges $\delta=\beta R_0$. This height difference may be compared to $\Delta h\sim \sigma H/E$ -- the approximate change in the sample’s height due to the compressive normal stress $(\sigma=F/\pi R_0^2)$. 
To eliminate the sensitivity of the delamination process to the tilt angle, we require $\delta\ll\Delta h$, which can be divided by $R_0$ to form a requirement on the tilt angle: $\beta\ll \beta_c= (180/\pi)\Delta h/R_0$ (in degrees). After substituting the values of the different quantities, for our smaller samples we have $\beta_c\sim0.36$ degrees, while for the larger samples $\beta_c\sim 0.11$ degrees.  Eliminating small angles of the order of less than 0.1 degrees is prohibitive in our system, thus explaining the difference in imperfection sensitivity among our observations and validating our decision to miniaturize the system for our main analysis. 

\section{PDMS Sample Preparation}
\label{appendix:PDMSSamplePreparation}
\setcounter{figure}{0} 

Our Sylgard 184 polydimethylsiloxane (PDMS) samples are fabricated using a protocol modified from \citet{raayai2019volume,chockalingam2021probing}, with a base:crosslinker ratio of 40:1. The cylindrical molds are of radius 1.198 cm, and the sample heights, measured after removal from the molds, were between 1.0 cm and 1.1 cm. The PDMS samples were fabricated in the laboratory according to the following protocol:

\smallskip

\noindent \underline{\textit{Step 1}:} We measured the base and cross-linker in the specified ratio of 40:1, and we poured both into the same mixing cup.

\smallskip

\noindent \underline{\textit{Step 2}:} We sealed the cup and mixed its contents using the default settings of a Thinky ARE-310 laboratory mixer. The default protocol mixed the contents for 30 seconds, then the machine paused, then the contents were mixed for an additional 30 seconds.

\smallskip
\noindent \underline{\textit{Step 3}:} We sprayed Pol-Ease 2300 Release Agent into the interior of each mold (cylindrical molds with a radius of 1.198 cm and a height of 1 cm) until it coated all surfaces of the mold. Since the molds were silicone-based like PDMS, the presence of the barrier layer from the mold release agent prevented the PDMS from bonding to the molds as it cured, which allowed for easier extraction of the samples from the molds after curing.

\smallskip
\noindent \underline{\textit{Step 4}:} Once the mold release agent dried, we poured the PDMS mixture into the molds until the molds were full, and we wiped excess PDMS off the molds.

\smallskip
\noindent \underline{\textit{Step 5}:} We placed the molds in a vacuum chamber and degassed them until bubbles stopped appearing on the top surface. Throughout the degassing process, we periodically released the vacuum, opened the vacuum chamber, and used a needle to pop the bubbles that had risen to the surface, to prevent the molds from overflowing.

\smallskip
\noindent \underline{\textit{Step 6}:} We allowed the samples to cure at room temperature for 1 hour.

\smallskip
\noindent \underline{\textit{Step 7}:} We placed the samples in an oven at 100 degrees Celsius for 1 hour.

\smallskip
\noindent \underline{\textit{Step 8}:} We removed the samples from the molds and placed them on an aluminum foil sheet.

\smallskip
\noindent \underline{\textit{Step 9}:} We allowed the samples to sit for 1 week prior to running any experiments on them.

\begin{figure}[h]
    \centering
    \includegraphics[width=0.3\textwidth]{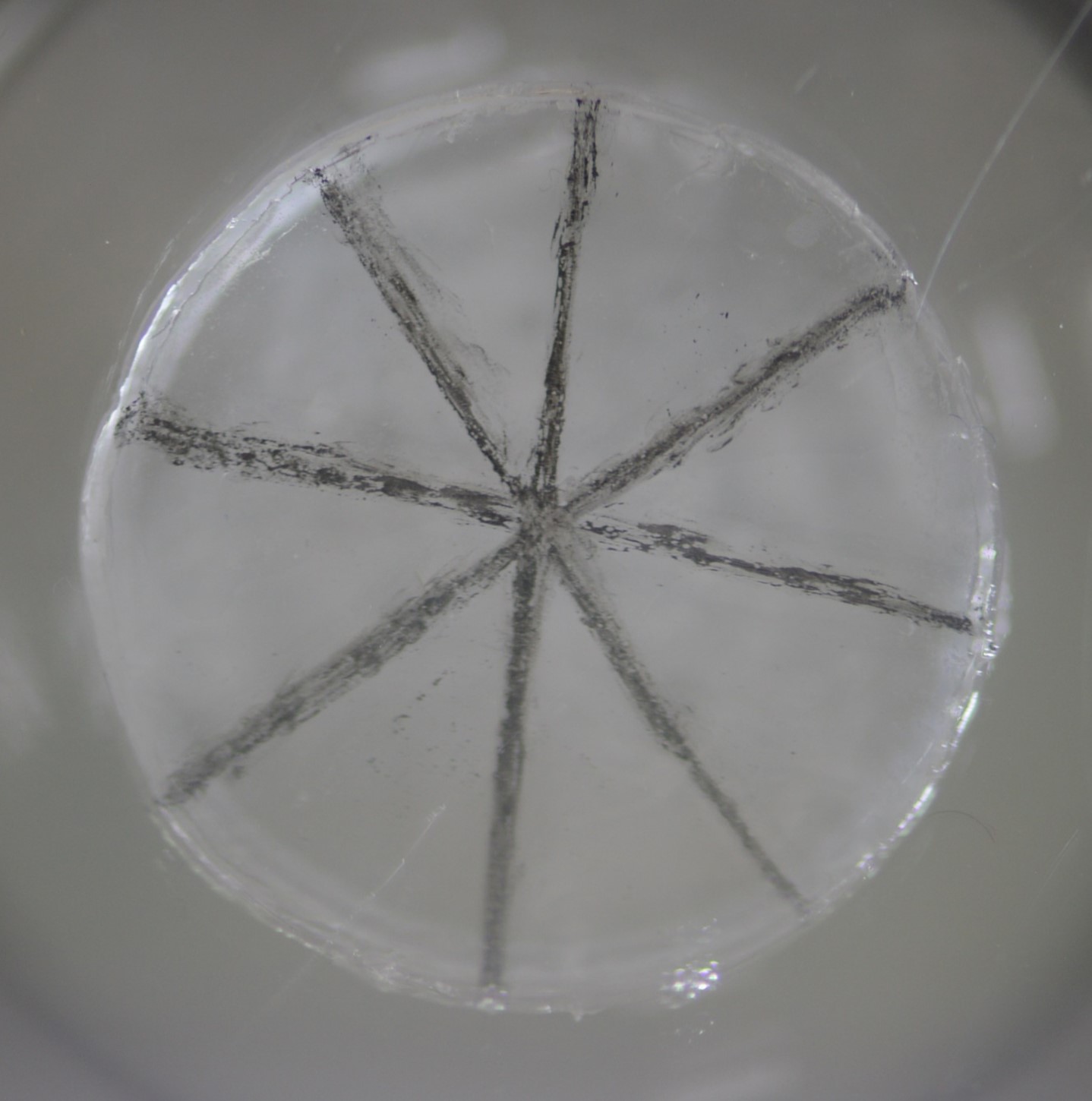} 
    \captionsetup{justification=centering}
    \caption{Radial markings on one circular face of SN125. The markings were initially equally-spaced straight lines, which allowed for easy tracking of their positions to monitor the delamination process.}
    \label{fig:RadialLines}
\end{figure}

\smallskip
\noindent \underline{\textit{Step 10}:} We marked 4 thin, equally spaced lines across the diameter on one face of each sample (on the circular face that had not come into contact with the mold release agent, so that any Pol-Ease residue would not affect the adhesive contact between the PDMS and the glass plate) in the pattern shown in Fig. \ref{fig:RadialLines}, using Pigma Micron 005 Archival Ink markers. The circular face with the radial lines was the face to be placed in contact with the glass plate for the experiments. 

\smallskip
\noindent \underline{\textit{Step 11}:} We adhered the circular face without markings to an acrylic disk via cyanoacrylate glue, which has shown to be effective in rigidly adhering PDMS to various substrates \citep{faley2015robust}. At this point, the samples were ready to be tested on the Instron after a 24-hour waiting period for the cyanoacrylate to cure.

\section{Experimental Setup and Protocol}
\label{appendix:ExperimentalSetupAndProtocol}
\setcounter{figure}{0} 

\begin{figure*}[h]
    \centering
    \includegraphics[width=0.6\textwidth]{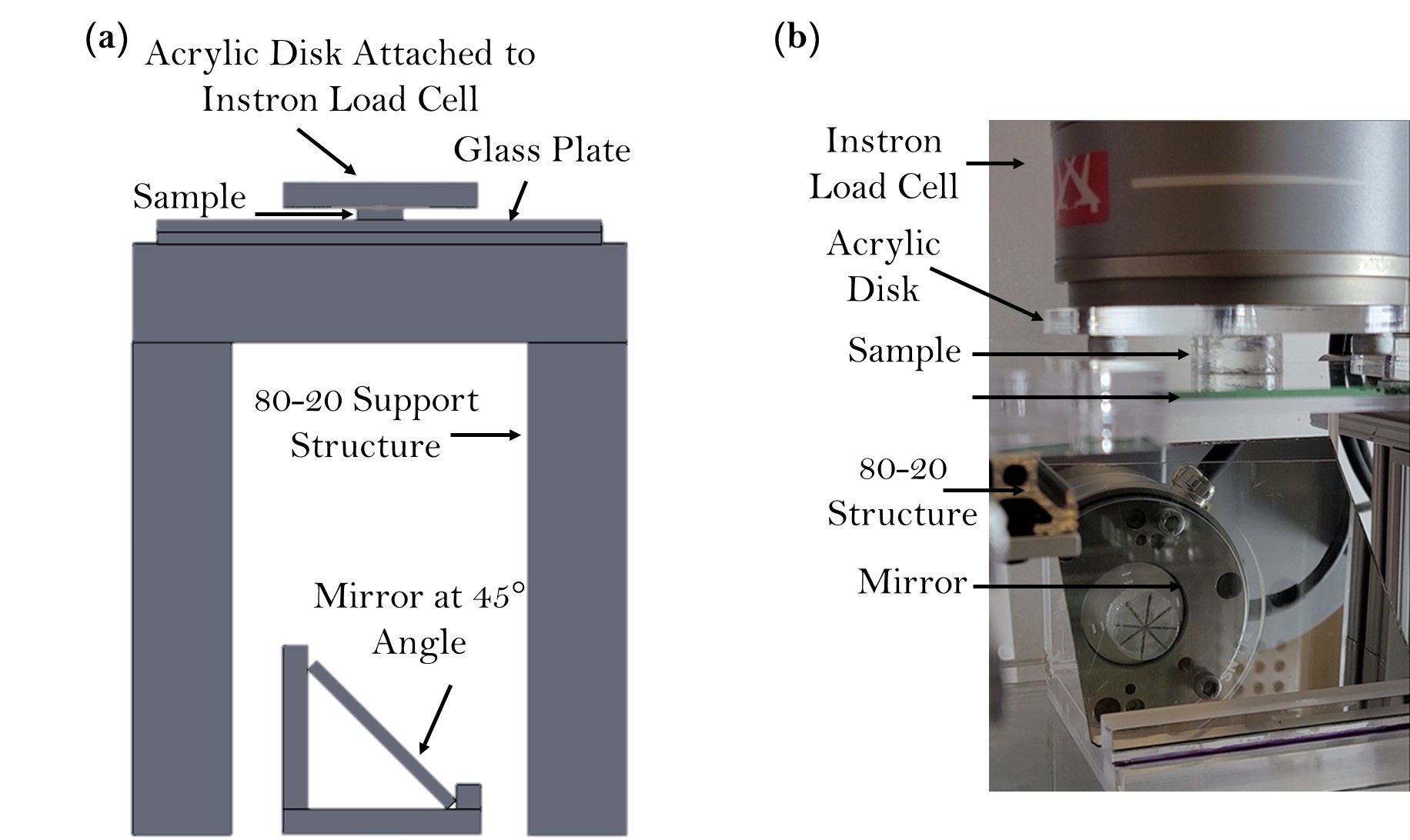} 
    \captionsetup{justification=centering}
    \caption{Experimental test setup: (a) Schematic and (b) image.}
    \label{fig:AppendixTestSetup}
\end{figure*}

The experimental test setup is shown as a schematic and as a labeled image in Fig. \ref{fig:AppendixTestSetup}. The PDMS sample was sandwiched between an acrylic disk (which is attached to the Instron load cell) and a glass plate. The side containing radial lines (discussed in \ref{appendix:PDMSSamplePreparation}) was in adhesive contact with the glass plate and was free to delaminate if necessary to achieve a state of lower energy. The curvature and movement of the initially-straight, equally-spaced lines were used to track the progression of delamination.

The side not containing the marked lines was rigidly adhered to the acrylic disk via cyanoacrylate glue, such that the top surface of the PDMS sample moved as a rigid body with the Instron load cell. The acrylic disks were etched with lines at regular intervals denoting the distance from the center of the disk, to ensure that the sample was centered on the disk when it was glued down (so that the torque from the Instron load cell would be applied through the central axis of the cylindrical sample). Each acrylic disk was given a Serial Number beginning with “SN” and followed by a 3-digit number, for easy identification during post-processing and analysis.

The glass plate was held above the laboratory bench by a series of 80-20 bars comprising a support structure. Underneath the bars, a mirror mounted at a 45-degree angle allowed for photography of the delamination of the bottom of the PDMS sample by placing a camera aimed horizontally at the mirror.

Prior to beginning the experimental tests, we turned on the Instron universal testing machine and allowed the sensors to stabilize for 15 minutes. We bolted the 80-20 support structure to the base plate of the Instron and also bolted the acrylic disk to the Instron load cell, then placed the angled mirror below the glass plate. We adjusted the height of the Instron crosshead until the bottom of the PDMS sample could make contact with the glass plate within the 6-cm vertical motion range of the load cell.

We placed a Nikon D3200 camera with its lens facing horizontally such that it could photograph the mirror image of the bottom of the PDMS sample. We set it to take images every 10 seconds for the duration of the test. We also positioned a Samsung Galaxy Note20 cell phone camera such that it could videotape the test, with both the lateral face of the cylinder and the mirror image of the bottom surface included in the frame of the video.

We calibrated the force and torque sensors and adjusted the vertical and rotary positions of the load cell such that it could carry out the vertical motion (to apply a normal force) and the 150-degree rotation without reaching the limits of the Instron’s motion. We then set the limits of the test to be outside the predicted range of motion of the test and armed them.

Using the Instron software, we programmed the Instron to apply a fixed initial compressive normal force (0 N, 0.5 N, or 1 N) to the sample before starting the rotation. We then allowed the torque and normal force to vary as necessary to provide the linear hold and rotary displacement required for the test. The Instron provided measurements of the torque and normal force every 0.2 seconds; we constructed a moving mean every 100 datapoints to reduce noise (Fig. \ref{fig:TorqueAndNormalForce} in the main text). Details of the program implementation are given below:\\


\noindent \underline{\textit{Step 1}:} The test was allowed to sit for 10 seconds to equalize and obtain an initial reading of the initial linear and angular position values.

\smallskip
\noindent \underline{\textit{Step 2}:} The Instron load cell was lowered until the normal force reached 1.2 N, where the positive value indicates a compressive normal force (the desired compressive normal force was 1 N, but given the noise in the load cell’s reading of the normal force, 1.2 N was chosen as the desired peak value so that the average reading would be approximately 1 N). This step ensured that the bottom surface of the PDMS was fully adhered to the glass plate.

\smallskip
\noindent \underline{\textit{Step 3}:} For tests that used 0 N (SN113 through SN118) or 0.5 N (SN119 through SN122) of initial compression, the Instron load cell was raised until the normal force peaked at -0.2 N or 0.3 N, respectively (again to account for noise in the data). This step was skipped for tests that used 1 N of initial compression (SN123 through SN126).

\smallskip
\noindent \underline{\textit{Step 4}:} After the desired normal force value was reached, the height of the load cell was held constant, and the rotation began. The load cell rotated at a constant rate of 0.2 degrees/second until it had reached 150 degrees of rotation relative to the starting point. During this rotation, the torque and normal force exerted by the load cell on the sample were graphed so that patterns could be observed.\\

After the test program completed, we removed the sample from the Instron by raising the height of the load cell until the PDMS sample was no longer in contact with the glass plate, then removing the bolts securing the acrylic disk to the Instron load cell. We examined the sample to determine whether fractures had formed on the lateral surface, near the face adhered to the acrylic disk. We then converted the raw data collected by the Instron WaveMatrix software and the two cameras into the results given in Section \ref{section:ExperimentalObservations}. Based on the torque and normal force trends and the amount of fracturing, we identified three outlier tests: SN117 (0 N), SN121 (0.5 N), SN126 (1 N). These tests are not included in the Section \ref{section:ExperimentalObservations} results.

\section{Experimental Observations of Wrinkling and Fracturing Patterns}
\label{appendix:WrinklingAndFracturing}
\setcounter{figure}{0} 

\begin{figure}[H]
    \centering
    \includegraphics[width=0.45\textwidth]{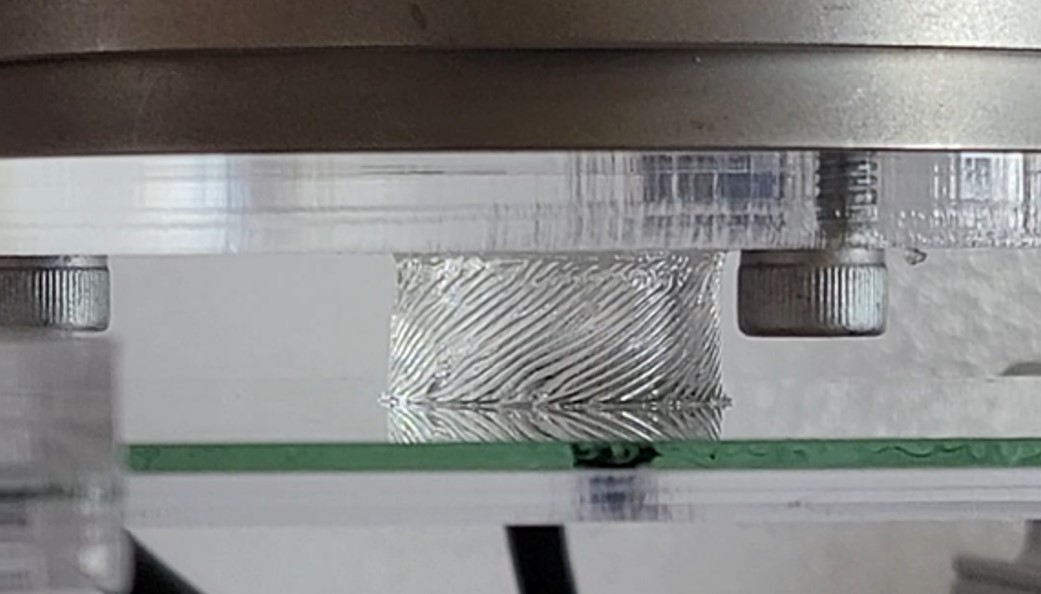} 
    \captionsetup{justification=centering}
    \caption{Wrinkles on the lateral surface of SN113 (0 N) at 89 degrees rotation. The wrinkles are at an oblique angle in the deformed configuration and extend throughout the height of the lateral surface.}
    \label{fig:Wrinkles}
\end{figure}

In addition to the delamination patterns discussed in the main text, we also observed wrinkles and fractures during the tests. Wrinkles formed at an oblique angle on the lateral surface of the samples, generally starting before 15 degrees of rotation. The wrinkles began near the bottom of the lateral surface and progressed upward along the lateral surface of the cylinder (Fig. \ref{fig:Wrinkles}). One possible explanation for this observation is that, because the outer edge of the cylinder experienced the maximum relative rotation between the top and bottom surface compared to any other region of the sample, the high strain on the lateral surface induced the wrinkling instability. This behavior experimentally validates the theoretical model developed by \citet{ciarletta2014torsion}, where torsion of a short cylinder leads to the formation of wrinkles on the lateral surface. Note that the type of instability experienced by a cylinder under torsion is highly dependent on the dimensions of the cylinder -- for long, thin cylinders, a knot-like instability will form \citep{gent2004torsional} which is not observed in our experimental set-up since the displacement along the axis of the cylinder is prevented by the geometrical constraints of our short and thick, "stubby" cylinder.

For the 0 N cases, the wrinkles seemed to disappear at high twist angles. Since delamination is easier to achieve in tests with no compressive normal force, perhaps delamination decreased the relative rotation between the top and bottom of these samples (and therefore the strain on the lateral surface) enough that the wrinkles disappeared.

Fracturing typically occurred toward the end of a given test. The fractures initiated on the lateral surface at the edge attached to the acrylic disk, then propagated downward as the rotation continued. The tests with no applied compression tended not to have any fractures (all the strain energy was relieved by delamination instead of fracturing). The 0.5 N cases tended to have small fractures, while the 1 N cases often had large fractures that ran through the thickness of the sample. 

\begin{figure}[H]
    \centering
    \includegraphics[width=0.45\textwidth]{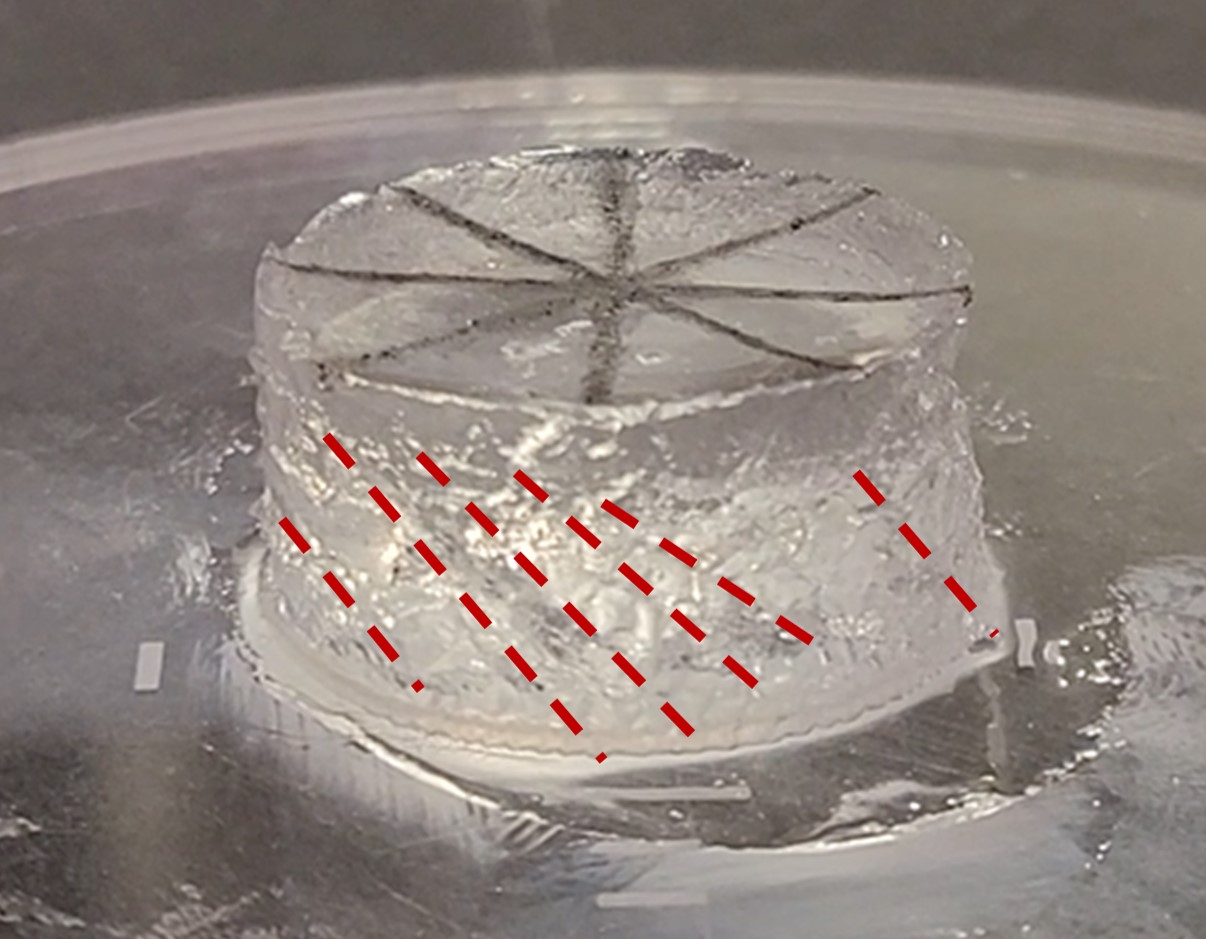} 
    \captionsetup{justification=centering}
    \caption{Image of the fractures on SN123 (1 N) after removal from the Instron. The fractures appear at an oblique angle in the undeformed (reference) configuration. Consistent with the results of the other 1 N tests, the fractures permeate the majority of the thickness of the sample, originating at the end of the lateral surface attached to the acrylic disk.}
    \label{fig:Fractures}
\end{figure}

In the images of the bottom surface of the PDMS, the fractures appeared at high angles of rotation as opaque white regions in the background near the outer edges of the circles (see the \href{https://drive.google.com/drive/u/0/folders/1XjKZ8HKTsY8BhSIHX8Wy9IqrMPHnHHpo}{videos}). In the videos of the lateral surface of the samples, the fractures formed vertically in the deformed configuration. After the samples were removed from the Instron, the fractures were oriented at an oblique angle in the undeformed (reference) configuration, somewhat similar to the angle at which wrinkles had appeared in the deformed configuration. An example of the fractures, photographed after the sample had been removed from the Instron, is shown in Fig. \ref{fig:Fractures}. However, our primary goal in this work was to study the delamination along the PDMS-glass interface, rather than cohesive failure of the PDMS itself. Thus, we did not examine the fracturing process in further detail.

\section{Curves of Delaminated Radius Overlaid with Torque and Normal Force Graphs}
\label{appendix:GraphsOfDelaminatedRadius}
\setcounter{figure}{0} 

\begin{figure*}[h]
    \centering
    \includegraphics[width=0.8\textwidth]{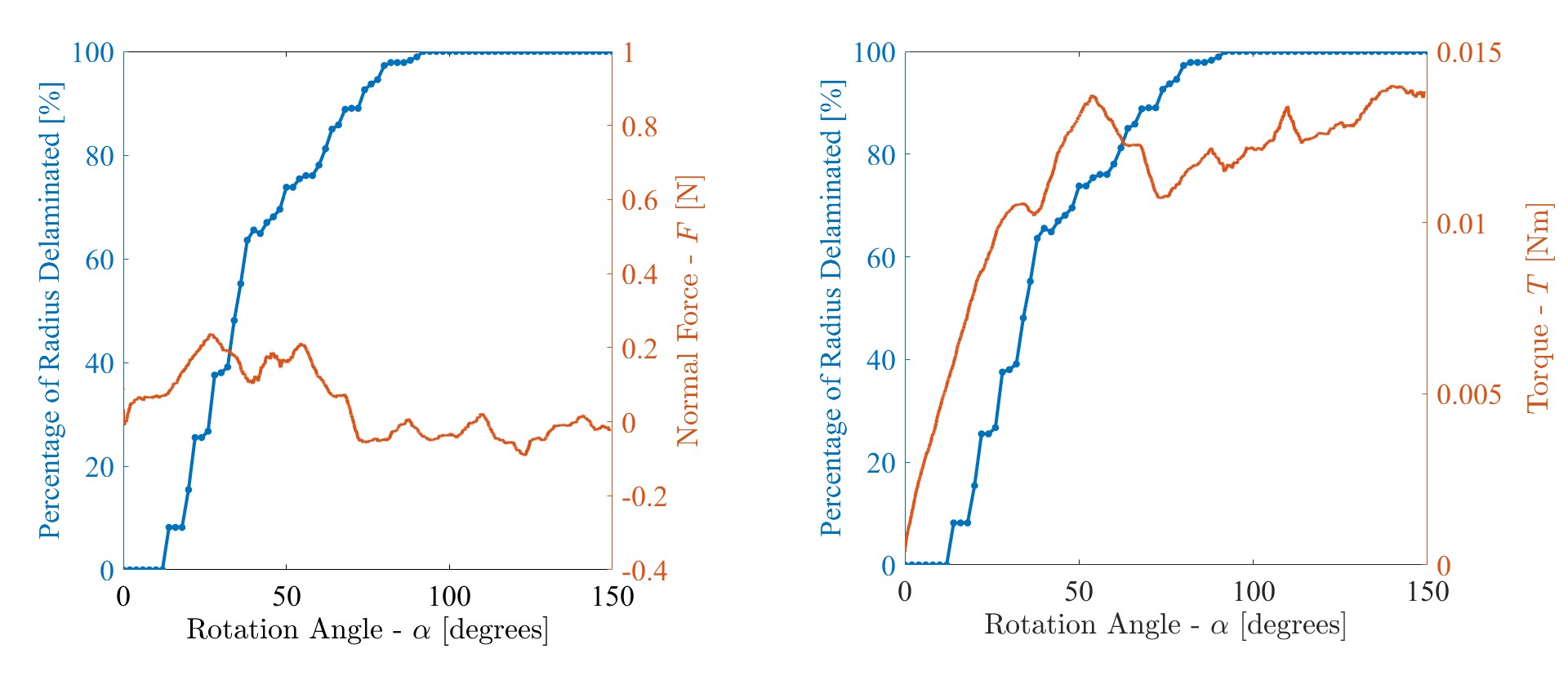} 
    \captionsetup{justification=centering}
    \caption{Graph of delaminated radius percentage overlaid on (left) normal force and (right) torque for SN114 (0 N). The normal force graph shows no discernible pattern, but the initial fluctuations in the torque correlate with steep jumps in the delaminated radius. Once the entire surface delaminates (at 92 degrees), the torque fluctuates randomly, governed by the stick-slip cycles that continue to propagate circumferentially throughout the radius.}
    \label{fig:0NAppendixGraphs}
\end{figure*}

\begin{figure*}[h]
    \centering
    \includegraphics[width=0.8\textwidth]{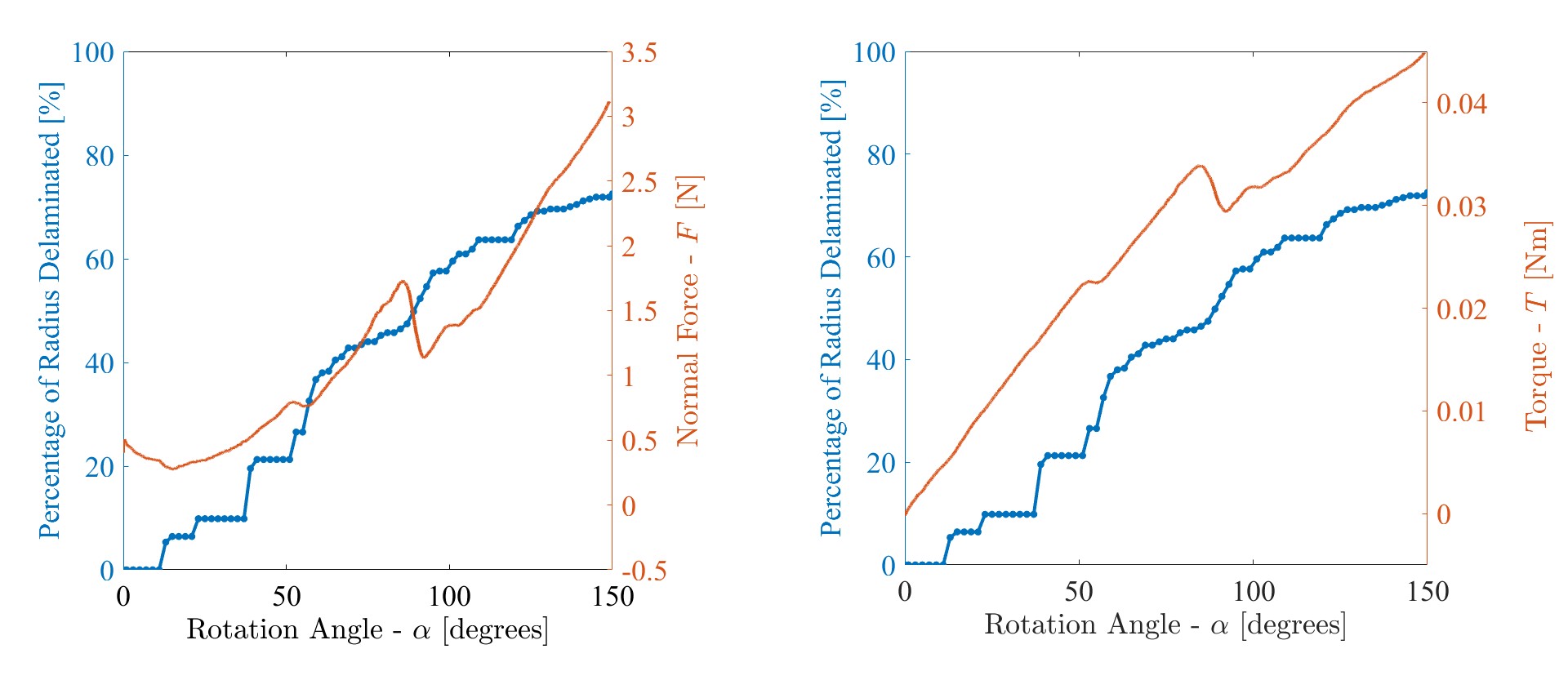} 
    \captionsetup{justification=centering}
    \caption{Graph of delaminated radius percentage overlaid on (left) normal force and (right) torque for SN119 (0.5 N). The steep jumps in delaminated radius correlate well with the plateaus and reversals in torque and normal force.}
    \label{fig:HalfNAppendixGraphs}
\end{figure*}

\begin{figure*}[h]
    \centering
    \includegraphics[width=0.8\textwidth]{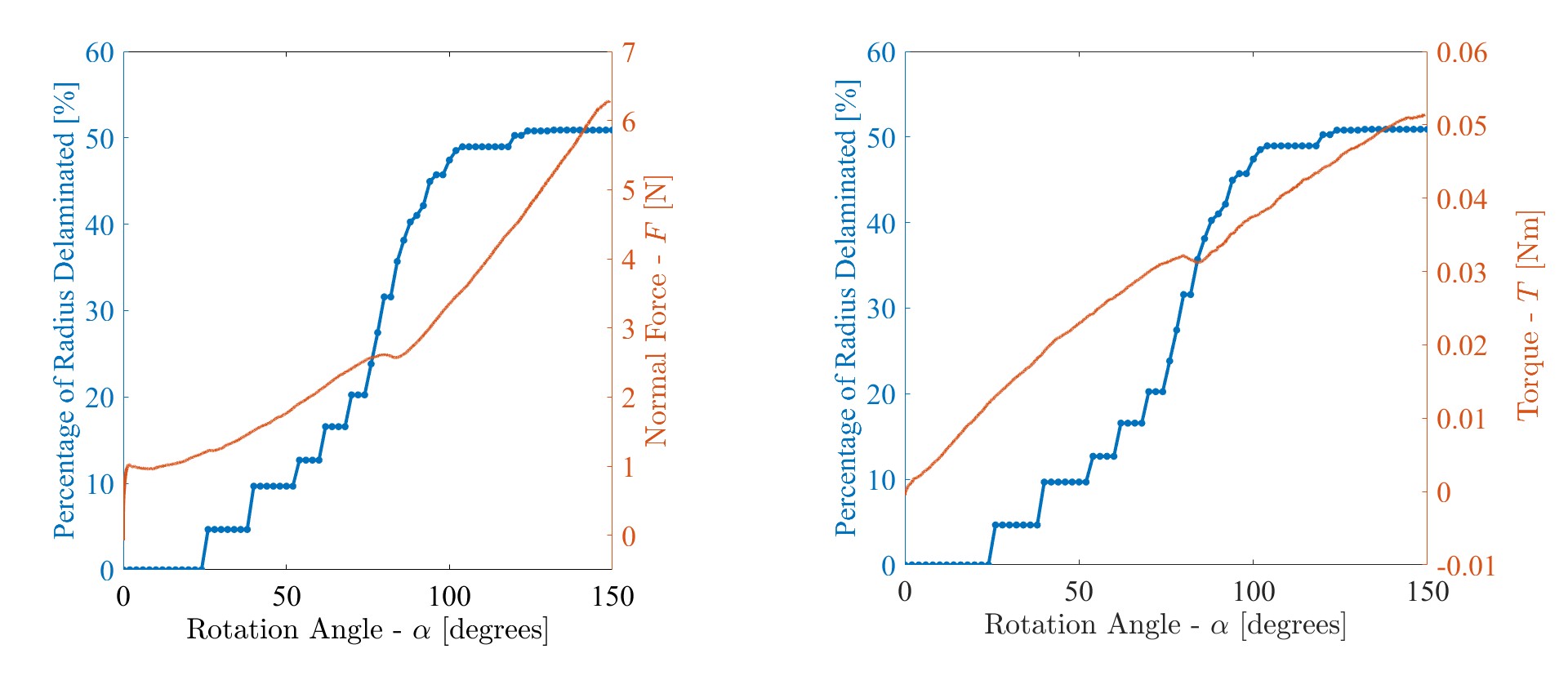} 
    \captionsetup{justification=centering}
    \caption{Graph of delaminated radius percentage overlaid on (left) normal force and (right) torque for SN123 (1 N). There is only one small reversal in torque and normal force, but it correlates with a jump in the delaminated radius.}
    \label{fig:1NAppendixGraphs}
\end{figure*}

The graphs below show the data from the representative tests in each normal force category: SN114 for 0 N, SN119 for 0.5 N, and SN123 for 1 N. The data of average delaminated radius percentage vs. rotation angle are overlaid on graphs of torque and normal force vs. rotation angle. The data presented here are the same as the graphs in Section \ref{section:ExperimentalObservations}, but the overlays provide insight into the correlation between the torque and normal force values and the progression of delamination. The overlays for SN114 are given in Fig. \ref{fig:0NAppendixGraphs}, the SN119 data are given in Fig. \ref{fig:HalfNAppendixGraphs}, and the SN123 results are given in Fig. \ref{fig:1NAppendixGraphs}.

Consistent with the results for most tests with no initial normal force, SN114 had many fluctuations in the torque and normal force values. The second half of the torque graph and the entire normal force graph in Fig. \ref{fig:0NAppendixGraphs} show so many abrupt reversals that no overarching trend could be observed. The initial fluctuations in torque (though not normal force, which shows no discernible pattern throughout the test) are correlated with steep jumps in the delaminated radius percentage. The entire bottom surface of the PDMS has delaminated by 92 degrees of rotation in this test, so the rapid fluctuations after that point (corresponding to stick-slip cycles moving circumferentially, causing the PDMS to delaminate and re-adhere) are to be expected. Thus, the results of the torque vs. rotation angle graph for SN114 may be explained by examining the delaminated radius plot, although the normal force graph does not exhibit the same trends. 

The results are clearer and require less detailed interpretation for SN119 and SN123. In Figs. \ref{fig:HalfNAppendixGraphs} and \ref{fig:1NAppendixGraphs}, the plateaus and reversals in both normal force and torque correlate well with the steep jumps in the delaminated radius. When the bottom surface delaminates, the difference between the rotation angle of the top surface and bottom surface of the PDMS decreases. Thus, the strain energy of the deformed PDMS sample is reduced, and the system requires less torque and compressive force to maintain the position of the load cell and acrylic, so the torque and normal force drop.

\end{document}